\documentclass[journal,draftcls,onecolumn,12pt,twoside]{IEEEtran}
\usepackage{graphicx}
\usepackage{amsmath, amsfonts, amssymb, amsthm}
\usepackage{enumerate}
\usepackage{multirow}
\usepackage{xcolor}
\theoremstyle{definition} \newtheorem{theorem}{Theorem}
\theoremstyle{definition} \newtheorem{corollary}[theorem]{Corollary}
\theoremstyle{definition} \newtheorem{proposition}[theorem]{Proposition}
\theoremstyle{definition} \newtheorem{definition}[theorem]{Definition}
\theoremstyle{definition} \newtheorem{lemma}[theorem]{Lemma}
\theoremstyle{definition} \newtheorem{algorithm}{Algorithm}

\theoremstyle{definition} \newtheorem*{convention}{Conventions}
\theoremstyle{definition} \newtheorem*{remark}{Remark}
\theoremstyle{definition} 
\newenvironment{Bullet}
{\begin{list}{}%
             {\setlength\labelsep{0pt}%
              \setlength\itemindent{0pt}%
              \setlength\leftmargin{18pt}%
              \setlength\labelwidth{18pt}%
              \setlength\topsep{0pt}%
              \setlength\parsep{0pt}%
              \setlength\itemsep{0pt}%
              }
\item[$\thickspace(*)\thickspace$\hfill]}%
{\end{list}}

\begin{document}
%

\sloppy

\title{On Base Field of Linear Network Coding}
\author{\IEEEauthorblockN{Qifu~Tyler~Sun,~Shuo-Yen~Robert~Li,~and~Zongpeng~Li}
\thanks{This work was presented in part at the 2015 International Symposium on Network Coding (Netcod).
}
}
\maketitle

\begin{abstract}
For a (single-source) multicast network, the size of a base field is the most known and studied algebraic identity that is involved in characterizing its linear solvability over the base field. In this paper, we design a new class $\mathcal{N}$ of multicast networks and obtain an explicit formula for the linear solvability of these networks, which involves the associated coset numbers of a multiplicative subgroup in a base field. The concise formula turns out to be the first that matches the topological structure of a multicast network and algebraic identities of a field other than size. It further facilitates us to unveil \emph{infinitely many} new multicast networks linearly solvable over GF($q$) but not over GF($q'$) with $q < q'$, based on a subgroup order criterion. In particular, i) for every $k\geq 2$, an instance in $\mathcal{N}$ can be found linearly solvable over GF($2^{2k}$) but \emph{not} over GF($2^{2k+1}$), and ii) for arbitrary distinct primes $p$ and $p'$, there are infinitely many $k$ and $k'$ such that an instance in $\mathcal{N}$ can be found linearly solvable over GF($p^k$) but \emph{not} over GF($p'^{k'}$) with $p^k < p'^{k'}$. On the other hand, the construction of $\mathcal{N}$ also leads to a new class of multicast networks with $\Theta(q^2)$ nodes and $\Theta(q^2)$ edges, where $q \geq 5$ is the minimum field size for linear solvability of the network.

\end{abstract}

\section{Introduction}
\label{Sec:Introduction}
Consider a communication network, which is a finite directed acyclic multigraph with each edge representing a noiseless transmission channel of unit capacity. There is a set of source nodes, each of which independently generates a set of data symbols belonging to a base field and transmits the data symbols simultaneously along the network. There is another set of receiver nodes, each of which attempts to recover a certain subset of source data symbols. A network is {\em linearly solvable} over a base field GF($q$) if there is a linear network coding scheme, which encodes every outgoing symbol of a node as a GF($q$)-linear combination of the incoming data symbols to this node, so that every receiver can recover its desired source data symbols at the same time.

\begin{figure}[htbp]
\centering
\scalebox{0.78}
{\includegraphics{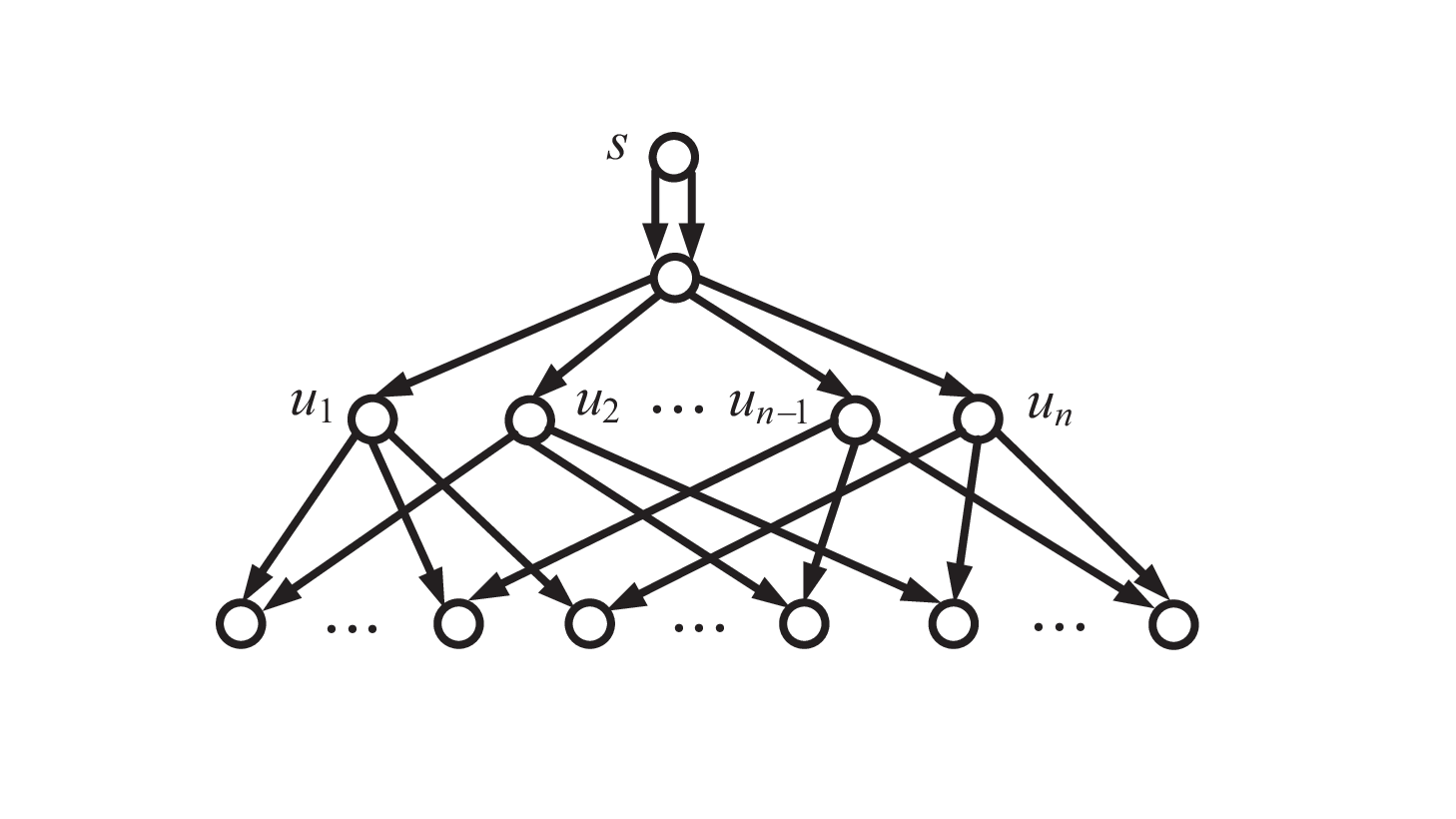}}
\caption{The $(n,2)$-combination network consists of nodes on 4-layers. The unique source $s$ generates two data symbols to be propagated to all bottom-layer nodes at the same time. The network is linearly solvable over GF($q$) if and only if $q \geq n - 1$, where $n$ is the topological parameter for the number of layer-3 nodes.}
\label{Fig:Combination_Network}
\end{figure}

The algebraic structure of a base field is closely related to the linear solvability of a network. When a network is (single-source) multicast, {\em i.e.}, every receiver attempts to recover the same set of data symbols transmitted from a unique source, the fundamental theorem of linear network coding \cite{LiYeungCai03} guarantees the existence of a linear solution when the base field is sufficiently large. Since the connection of field size on the linear solvability of a multicast network was revealed, there have been extensive studies on the field size requirement and efficient construction of a linear solution [2-12]. Among them, the best known explicit sufficient condition for a multicast network to be linearly solvable over GF($q$) is $q \geq |T|$, where $|T|$ is the number of receivers \cite{Jaggi05}\cite{Langberg09}. Moreover, the combination networks \cite{Ngai_Yeung}\cite{Xiao07} constitute the only investigated class of multicast networks with an explicit linear solvability characterization that matches algebraic identities of GF($q$) and topological parameters of the network. Specifically, an $(n, 2)$-combination network, as depicted in Fig. \ref{Fig:Combination_Network}, is known to be linearly solvable over GF($q$) if and only if $q \geq n - 1$, where $n$ is the topological parameter for the number of layer-3 nodes.

Without the restriction of considering a multicast network, more algebraic identities inherited in a field other than size have been revealed to affect the linear solvability of a network. In \cite{DFZeger05}, two networks were designed, one of which is linearly solvable only over a field with even characteristic, while the other is linearly solvable only over a field with odd characteristic. It is further unveiled in \cite{DFZeger08} that for an arbitrary finite or co-finite set $S$ of prime numbers, a network can be constructed such that the network is linearly solvable over some field of characteristic $p$ for every $p$ in $S$, whereas it is not linearly solvable over any field whose characteristic is not in $S$. Meanwhile, whether a set of polynomials with integer coefficients has common roots is connected to a network linear solvability problem. Specifically, a general method was proposed in \cite{DFZeger08} to associate a polynomial set with a network, such that the polynomials have common roots over a field if and only if the corresponding network has a linear solution over the same field. In \cite{Dey}, analogous equivalence was also established between the existence of common roots over a finite field for a polynomial set and the existence of a linear solution over the same field for an associated {\em sum-network}, in which all receivers attempt to recover the sum of all source data symbols. However, when attention is only paid to multicast networks, these results are not applicable anymore. First, since a multicast network is linearly solvable over a large enough field, it is linearly solvable over some field of characteristic $p$ for every prime $p$. Moreover, the general method proposed in \cite{DFZeger08} and \cite{Dey} cannot necessarily construct a multicast network from a polynomial set.

If the field size were the only algebraic factor that affects the linear solvability of a multicast network, then all multicast networks would conceivably share a property $q^*_{max} < q_{min}$, where $q_{min}$ refers to the minimum field size for the existence of a linear solution and $q^*_{max}$ refers to the maximum field size for the non-existence of a linear solution ($q^*_{max}$ is set to 1 if the network is linearly solvable over every field.) The first few exemplifying networks with the special property $q^*_{max} > q_{min}$ were not discovered until a rather recent work \cite{Sun_ISIT14}. As pointed out in \cite{Sun_ISIT14}, there are fundamental reasons underlying the intriguing observation that $q^*_{max} > q_{min}$ is possible for multicast networks: not only the field size $q$, but also \emph{the order of proper subgroups in the multiplicative group} GF($q$)$^\times$ = GF($q$)$\backslash \{0\}$ of GF($q$) plays an important role in the linear solvability over GF($q$). However, even though it was further observed in \cite{Sun_TIT} that all these newly designed networks with $q^*_{max} > q_{min}$ share a common topological structure, the intrinsic impact of multiplicative subgroup orders of a field on the linear solvability of multicast networks remains unclear.

In this paper, we explicate in depth how multiplicative subgroups in GF($q$) can affect the linear solvability of multicast networks over GF($q$), as outlined as follows.

\begin{itemize}
\item In Sec. \ref{Sec:General_Framework}, we construct a general class $\mathcal{N}$ of layered multicast networks, which subsumes all multicast networks presented in \cite{Sun_TIT} with $q_{min} < q^*_{max}$ as special instances. After deriving a generalized Cauchy-Davenport theorem, we proceed to construct an explicit formula for the linear solvability of networks in $\mathcal{N}$ over GF($q$). Besides the topological parameters of $\mathcal{N}$, the concise formula involves the associated coset numbers of a multiplicative subgroup in GF($q$). It turns out to be the first that matches the topological structure of a multicast network and algebraic identities of a field other than size.
\item In Sec. \ref{Sec:Subgroup_Order_Criterion}, based on the general characterization of $\mathcal{N}$, we further formulate a subgroup order criterion for a pair of finite fields. As long as (GF($q$), GF($q'$)) satisfies the subgroup order criterion, where $q < q'$ is possible, we can establish an instance in $\mathcal{N}$ that is linearly solvable over GF($q$) but not over GF($q'$).
\item In Sec. \ref{Sec:Establish_Instances}, as an application of the subgroup order criterion, we are able to establish \emph{infinitely many} multicast networks in $\mathcal{N}$ with $q_{min} < q^*_{max}$. As intriguing instances, for every $k \geq 2$, we can establish a multicast network with $q_{min} = 2^{2k}$ and $q^*_{max} = 2^{k'}$ for some $k' > 2k$. This proves that $q^*_{max} - q_{min} > 0$ can tend to infinity for a multicast network. Moreover, for \emph{arbitrary} distinct primes $p$ and $p'$, there are \emph{infinitely many} $k$ and $k'$ such that an instance in $\mathcal{N}$ can be found to be linearly solvable over GF($p^k$) but \emph{not} over GF($p'^{k'}$) with $p^k < p'^{k'}$.
\item In Sec. \ref{Sec:Construction_Network_Prescribed_q_min}, based on the general characterization of $\mathcal{N}$, a new procedure is proposed to construct a multicast network with $q_{min}$ equal to any prescribed prime power $q \geq 5$, and the constructed network has a smaller size compared with the $(q+1, 2)$-combination network, which has $q_{min} = q$ too.
\end{itemize}

To summarize, we systematically develop a framework to reflect the intrinsic impact of multiplicative subgroups on the linear solvability of a multicast network. A number of new results on the comparison between $q_{min}$ and $q^*_{max}$ are also subsequently deduced. Our findings suggest that a ``matching'' between the algebraic structure of a base field and the topological structure of a multicast network is necessary for the existence of a linear solution to the multicast network.

\vspace{5pt}

\section{A Role of Multiplicative Subgroups \\ on Linear Solvability of Multicast Networks}
\label{Sec:General_Framework}
\begin{convention}
A \emph{multicast network} is modeled as a finite directed acyclic multigraph with a unique source node $s$ and a set $T$ of receivers. The out-degree of $s$ is denoted by $\omega$ and is assumed equal to the source dimension, which means the number of data symbols to be simultaneously transmitted by the source. For a node $v$ in the network, denote by $In(v)$ the set of its incoming edges. For an arbitrary set $N$ of non-source nodes, denote by $maxflow(N)$ the maximum number of edge-disjoint paths starting from $s$ and ending at nodes in $N$. Every edge in the network is of unit capacity. A linear network code (LNC) over GF($q$) is an assignment of a coding coefficient $k_{d,e} \in \mathrm{GF}(q)$ to every pair $(d, e)$ of edges such that $k_{d,e} = 0$ when $(d, e)$ is not an adjacent pair. Every LNC uniquely determines a coding vector $f_e$, which is an $\omega$-dimensional column vector over GF($q$), for each edge $e$ in the network. A multicast network is linearly solvable over GF($q$) if there is an LNC over GF($q$) such that for each receiver $t \in T$, the $\omega \times |In(t)|$ matrix $[f_e]_{e\in In(t)}$ over GF($q$) is of full rank $\omega$. Such an LNC is called a linear solution over GF($q$) for the multicast network. Let $q_{min}$ be the minimum field size for the existence of a linear solution over GF($q_{min}$), and $q^*_{max}$ the maximum field size for the nonexistence of a linear solution over GF($q^*_{max}$). We set $q^*_{max}$ to $1$ if the network is linearly solvable over all finite fields. \end{convention}

\begin{figure}[htbp]
\centering
\scalebox{0.5}
{\includegraphics{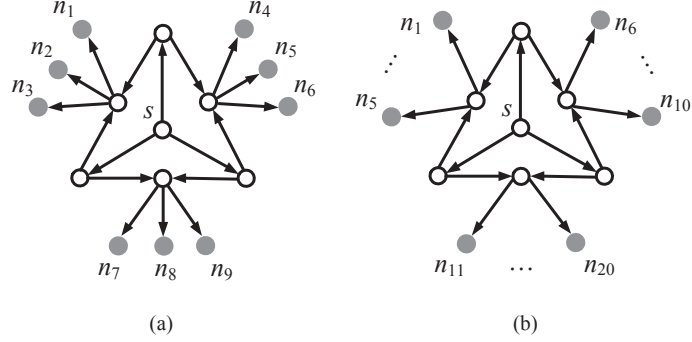}}
\caption{For both multicast networks, the source dimension $\omega$ is 3. For every set $N$ of 3 grey nodes that has $maxflow(N) = 3$, there is a receiver connected from it, which is omitted in the depiction for simplicity. The network (a) has $7 = q_{min} < q^*_{max} = 8$ and the network (b) has $16 = q_{min} < q^*_{max} = 17$.}
\label{Fig:Rank_3_networks}
\end{figure}

Fig. \ref{Fig:Rank_3_networks} reproduces two of the multicast networks discovered in \cite{Sun_ISIT14}, with $q_{min} < q^*_{max}$. In both networks, the source dimension $\omega$ is equal to $3$ and there is a non-depicted receiver connected from every set $N$ of three grey nodes whenever the maximum flow from the source to the nodes in $N$ is $3$. For example, there is a receiver connected from $\{n_1, n_2, n_7\}$ in Fig. \ref{Fig:Rank_3_networks}(a) and from $\{n_1, n_6, n_{11}\}$ in Fig. \ref{Fig:Rank_3_networks}(b). %
It can be shown that $q_{min} = 7$, $q^*_{max} = 8$ for the network in Fig. \ref{Fig:Rank_3_networks}(a) and $q_{min} = 16$, $q^*_{max} = 17$ for the network in Fig. \ref{Fig:Rank_3_networks}(b). In the course of characterizing the linear solvability of these networks, it has already been noted in \cite{Sun_TIT} that these exemplifying networks share a common topological structure, and a general 5-layer multicast network is correspondingly constructed and analyzed for unifying the proof. In this paper, we construct a more general class of multicast networks in the sense of involving more topological parameters as follows.

\begin{algorithm}
\label{Alg:General_Construction}
Given a positive integer $\omega \geq 3$ and an $\omega$-tuple $\mathbf{d} = (d_1, \cdots d_{\omega})$ of positive integers larger than 1 as input parameters, the procedure below constructs a multicast network $\mathcal{N}_{\omega,\mathbf{d}}$ with source dimension $\omega$ consisting of nodes on five layers, which are labeled 1-5 from upstream to downstream, and all edges are between adjacent layers.

\vspace{5pt}

\noindent \underline{\emph{Step 1}}. Create a source $s$, which forms the unique node at layer 1.

\vspace{3pt}

\noindent \underline{\emph{Step 2}}. Create $\omega$ layer-2 nodes, each of which is connected with $s$ by a single edge. Sequentially label these nodes as $u_1, u_2, \cdots, u_\omega$.

\vspace{3pt}

\noindent \underline{\emph{Step 3}}. Create $\omega$ layer-3 nodes, labeled as $v_1, v_2, \cdots, v_\omega$. Each node $v_i$, $1 \leq i \leq \omega$, has 2 incoming edges, one leading from $u_i$ and the other from $u_{i-1}$, where $u_0$ will represent $u_\omega$.

\vspace{3pt}

\noindent \underline{\emph{Step 4}}. For each layer-3 node $v_i$, $1 \leq i \leq \omega$, create $d_i$ downstream layer-4 nodes $n_{i, 1}, n_{i, 2}, \cdots, n_{i, d_i}$, each of which has the unique incoming edge $e_{ij}$ leading from $v_{i}$.

\vspace{3pt}

\noindent \underline{\emph{Step 5}}. For every set $N$ of $\omega$ layer-4 nodes with $maxflow(N) = \omega$, create a layer-5 node connected from every node in $N$ by an edge. Set all layer-5 nodes to be receivers.
\hfill $\blacksquare$
\end{algorithm}

\begin{figure}[t]
\centering
\scalebox{0.58}
{\includegraphics{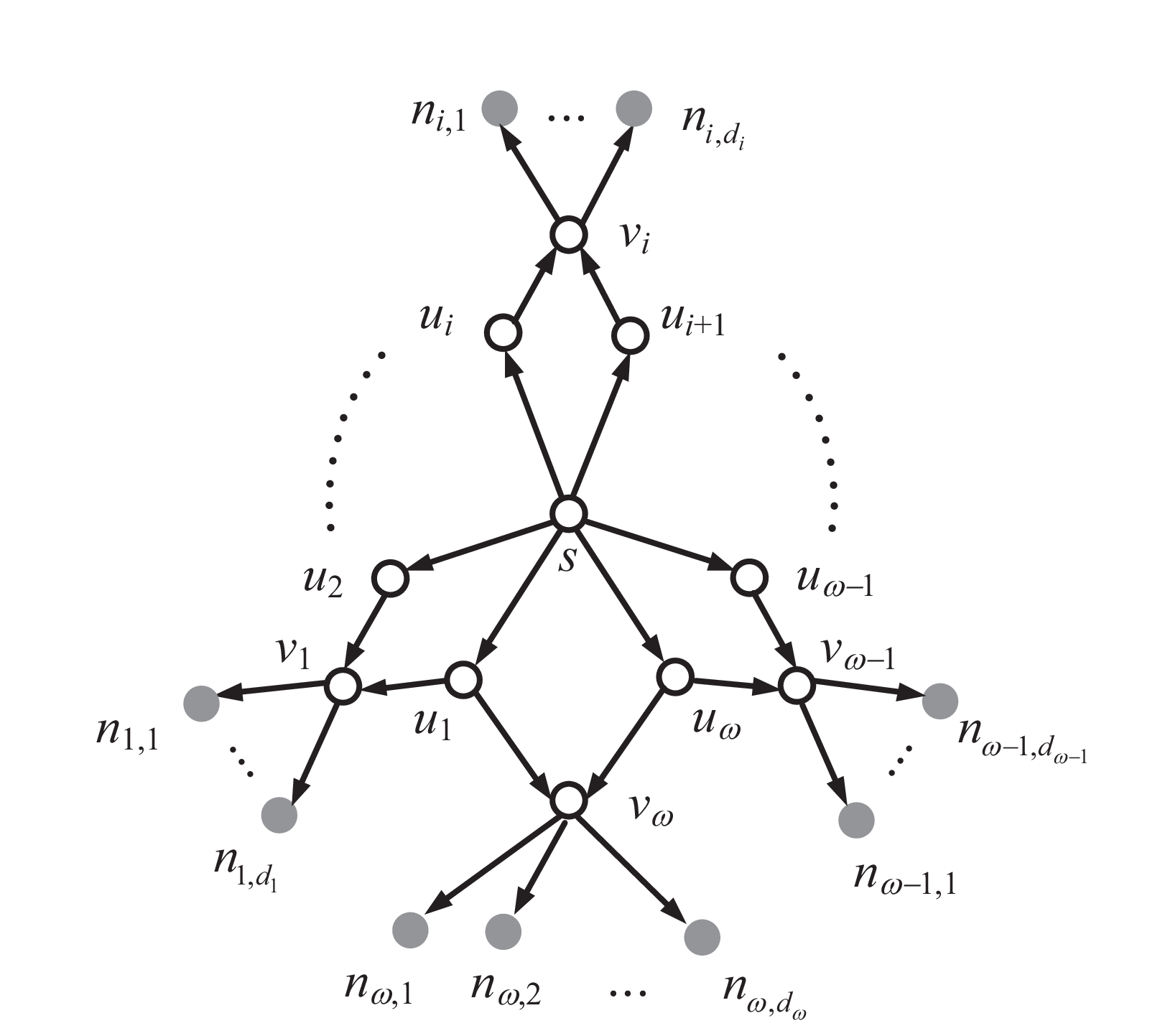}}
\caption{The general network $\mathcal{N}_{\omega, \mathbf{d}}$ constructed by Algorithm \ref{Alg:General_Construction} based on parameters $\omega$ and $\mathbf{d} = (d_1, \cdots, d_\omega)$ consists of nodes on 5 layers. The layer-1 just consists of the source node $s$, and all layer-4 nodes are depicted in grey. There is a non-depicted bottom-layer node connected from every set $N$ of $\omega$ layer-4 nodes with $maxflow(N) = \omega$. All bottom-layer nodes are receivers.}
\label{Fig:General_network}
\end{figure}

Fig. \ref{Fig:General_network} depicts the topology of a general $\mathcal{N}_{\omega, \mathbf{d}}$. The network in Fig. \ref{Fig:Rank_3_networks} can be constructed by Algorithm \ref{Alg:General_Construction} with respective parameters $\omega = 3, \mathbf{d} = (3, 3, 3)$ and $\omega = 3, \mathbf{d} = (5, 5, 10)$. Moreover, the general network considered in \cite{Sun_TIT} with parameters $\omega, d_1, d_2$ can also be regarded as a particular instance of $\mathcal{N}_{\omega,\mathbf{d}}$ with $\mathbf{d} = (\underbrace{d_1, d_1, \cdots, d_1}_{\omega-1}, d_2)$. The following equivalent conditions for the linear solvability of network $\mathcal{N}_{\omega,\mathbf{d}}$ can be readily derived.

\begin{lemma}
\label{lemma:general_network_linear_solvability}
Consider a network $\mathcal{N}_{\omega,\mathbf{d}}$ constructed by Algorithm \ref{Alg:General_Construction}. The followings are equivalent.
\begin{enumerate}[a)]
\item The network is linearly solvable over a finite field GF($q$).
\item There is a matrix completion of the matrix $\mathbf{M}$ depicted in (\ref{eqn:general_matrix_M_for_linear_solvability_analysis}) over GF($q$), that is, an assignment of values $a_{ij}$ in GF($q$) to all indeterminates $x_{ij}$ such that the rank of every $\omega\times \omega$ submatrix in $\mathbf{M}$ is preserved.

\begin{equation}
\label{eqn:general_matrix_M_for_linear_solvability_analysis}
\mathbf{M} = \left[
  \begin{array}{ccccccccccccc}
    1 & \cdots & 1 & 0 & \cdots & 0 &  & 0 & \cdots & 0 & x_{\omega1} & \cdots  & x_{\omega d_\omega} \\
    x_{11} & \cdots & x_{1d_1} & 1 & \cdots & 1 &  & \vdots & \ddots & \vdots & 0 & \cdots  & 0\\
    0 & \cdots & 0 & x_{21} & \cdots & x_{2d_2} &  & 0 & \cdots & 0 & 0 & \cdots & 0\\
    0 & \cdots & 0 & 0 & \cdots & 0 & \ddots & 0 & \cdots & 0 & \vdots & \ddots & \vdots \\
    \vdots & \ddots & \vdots & \vdots & \ddots & \vdots & & 1 & \cdots & 1 & 0 & \cdots & 0 \\
    0 & \cdots & 0 &  0 & \cdots & 0 &  & x_{(\omega-1)1} & \cdots & x_{(\omega-1)d_{\omega-1}} & 1 & \cdots & 1 \\
  \end{array}
\right]
\end{equation}
\item There exists a set $S_i = \{a_{i1}, \cdots, a_{id_i}\}$ of distinct nonzero elements in GF($q$) for each $1 \leq i \leq \omega$ such that
\begin{equation}
(-1)^{\omega-1} \notin S_1\cdot S_2 \cdot \cdots \cdot S_\omega = \{a_{1j_1}a_{2j_2}\cdots a_{\omega j_\omega}: 1 \leq j_k \leq d_k, 1 \leq k \leq \omega\}
\label{eqn:network_linear_solvability_rule}
\end{equation}
\end{enumerate}
Moreover, when condition c) holds for some $\{S_i\}_{1\leq i \leq \omega}$, the assignment of $x_{ij} = a_{ij}$ is a matrix completion of matrix $\mathbf{M}$, and a linear solution is given by $\mathbf{M}$ which represents the juxtaposed matrix of coding vectors for edges into layer-4 (grey) nodes.
\begin{proof}
Please refer to Appendix\ref{Appendix_general_linear_solvability}.
\end{proof}
\end{lemma}

The equivalent conditions in Lemma \ref{lemma:general_network_linear_solvability} implicitly match the topological parameters $\omega$ and $\mathbf{d}$ in $\mathcal{N}_{\omega, \mathbf{d}}$ with GF($q$), while the inherent algebraic identity that affects the linear solvability of $\mathcal{N}_{\omega,\mathbf{d}}$ remains unveiled. In the remaining part of this section, we shall proceed to derive an explicit equivalent condition that matches the parameters $\omega$ and $\mathbf{d}$ with not only the size, but also the multiplicative subgroup orders of GF($q$).

Recall that all nonzero elements in GF($q$) can be represented as $\xi^j$, $1 \leq j \leq q - 1$ for some element $\xi$. Such an element $\xi$ is called a \emph{primitive element} in GF($q$). Thus, $\mathrm{GF}(q)\backslash\{0\} = \{\xi, \xi^2, \cdots, \xi^{q-1}\}$. Since $\xi^{q-1} = \xi^0 = 1$, the set $\mathrm{GF}(q)\backslash\{0\}$ forms a cyclic group with generator $\xi$ and of order $q - 1$. This cyclic group is isomorphic to the additive group $\mathbb{Z}_{q-1}$ of integers modulo $q-1$ via the mapping of $\xi^k \mapsto k$. When the field GF($q$) has odd characteristic, $(-1)^{\omega-1} = \xi^{(\omega-1)(q-1)/2} \mapsto (\omega-1)(q-1)/2$. Else, $(-1)^{\omega-1} \mapsto 0$. Thus, the network $\mathcal{N}_{\omega, \mathbf{d}}$ has a linear solution over GF($q$) if and only if
\begin{itemize}
\item there exist subsets $T_1, T_2, \cdots, T_\omega$ of $\mathbb{Z}_{q-1}$ with respective cardinalities $d_1, d_2, \cdots , d_\omega$, such that the set $T_1 + T_2 + \cdots + T_\omega = \{b_1 + b_2 +\cdots + b_\omega: b_j \in T_j, 1 \leq j \leq \omega\}$ does not include $(q-1)/2$ when $q$ is odd and does not include $0$ when $q$ is even.
\end{itemize}
With a parallel shift on all members of $T_1$ if necessary, we have the following equivalent condition.

\begin{proposition}
\label{prop:linear_solvability_exhaust_claim}
The network $\mathcal{N}_{\omega, \mathbf{d}}$ has a linear solution over GF($q$) if and only if there exist subsets $T_1, T_2, \cdots, T_\omega$ of additive group $\mathbb{Z}_{q-1}$ with respective cardinalities $d_1, d_2, \cdots , d_\omega$ such that the set $T_1 + T_2 + \cdots + T_\omega$ does not exhaust $\mathbb{Z}_{q-1}$, that is,
\[
|T_1 + T_2 + \cdots + T_\omega| < q - 1.
\]
\end{proposition}

\vspace{6pt}
Consider the network $\mathcal{N}_{\omega, \mathbf{d}}$ with parameters $\mathbf{d} = (d_1, d_1, \cdots, d_1, d_2)$. If there is a positive integer $d > 1$ that divides $q-1$, then there is a subgroup $H$ of order $d$ in $\mathbb{Z}_{q-1}$. If $d_1 \leq d \leq q - 1 - d_2$, then we can assign each $T_j$, $1 \leq j \leq \omega -1$, to be arbitrary $d_1$ elements in $H$ and $T_\omega$ to be arbitrary $d_2$ elements in $\mathbb{Z}_{q-1} \backslash H$, \emph{i.e.}, the complement of $H$ in $\mathbb{Z}_{q-1}$. Thus, $T_1 + \cdots + T_{\omega-1} \subseteq H$ and $T_1 + \cdots + T_{\omega-1} + T_\omega = T_\omega \subseteq \mathbb{Z}_{q-1} \backslash H$. This implies that the set $T_1 + \cdots + T_\omega$ does not exhaust the whole $\mathbb{Z}_{q-1}$ and hence the network is linearly solvable over GF($q$). This sufficient condition in terms of subgroup orders was applied in \cite{Sun_TIT} for the purpose of proving the existence of linear solutions over GF($q$). For instance, for the network $\mathcal{N}_{\omega, \mathbf{d}}$ with $\mathbf{d} = (5, 5, 10)$ as depicted in Fig. \ref{Fig:Rank_3_networks}(b), it is linearly solvable over GF($16$) since we can set $T_1 = T_2 = \{0, 3, 6, 9, 12\}$ and $T_3 = \{1, 2, 4, 5, 7, 8, 10, 11\}$ in $\mathbb{Z}_{15}$ such that $T_1 + T_2 + T_3 = T_3 = \{1, 2, 4, 5, 7, 8, 10, 11\}$ does not exhaust $\mathbb{Z}_{15}$.  However, no handy necessary condition has been derived in \cite{Sun_TIT} for the nonexistence of linear solutions, and this is a key reason that only a few exemplifying networks with the special property $q_{min} < q^*_{max}$ were designed therein. For instance, the nonexistence of a linear solution over GF(17) for $\mathcal{N}_{\omega, \mathbf{d}}$ with $\mathbf{d} = (5, 5, 10)$ is merely verified in \cite{Sun_TIT} by exhaustive enumeration on all possible subsets of nonzero elements in GF($17$) for condition (\ref{eqn:network_linear_solvability_rule}) in Lemma \ref{lemma:general_network_linear_solvability}.

Recall that when $q-1$ is prime, Cauchy-Davenport Theorem asserts that for any two non-empty subsets $A$ and $B$ of $\mathbb{Z}_{q-1}$, $|A + B| \geq \min\{|A|+|B|-1, q-1\}$. Thus, for arbitrary three-element subsets $T_1, T_2, T_3$ of $\mathbb{Z}_7$, $|T_1 + T_2 + T_3| = 7$, that is, $T_1, T_2, T_3$ exhaust the whole $\mathbb{Z}_7$. Consequently, by applying Proposition \ref{prop:linear_solvability_exhaust_claim} to the network $\mathcal{N}_{3, \mathbf{d}}$ with $\mathbf{d} = (3, 3, 3)$, we conclude that the network depicted in Fig. \ref{Fig:Rank_3_networks}(a) is not linearly solvable over GF($8$).

We next generalize the Cauchy-Davenport Theorem to be over $\mathbb{Z}_{q-1}$, where $q-1$ can be a composite. This requires the concept of stabilizer group (See, e.g., \cite{Dummit_Foote}). Recall that the \emph{stabilizer group} of a subset $A$ in a group $G$ is defined as $\{g \in G: gA = A\}$.

\begin{theorem}
\label{thm:Cauchy-Davenport-Generalization}(Generalized Cauchy-Davenport Theorem over $\mathbb{Z}_{n}$)
Let $A_1, A_2, \cdots, A_k$ be nonempty subsets of $\mathbb{Z}_n$ and $H$ the stabilizer group of $A_1 + A_2 + \cdots + A_k$ in $\mathbb{Z}_n$. Then
\begin{equation}
\label{eqn:Kneser_Generalization}
|A_1 + A_2 + \cdots + A_k| \geq |H|\left(\left\lceil\frac{|A_1|}{|H|}\right\rceil + \left\lceil\frac{|A_2|}{|H|}\right\rceil + \cdots + \left\lceil\frac{|A_k|}{|H|}\right\rceil - (k-1)\right)
\end{equation}
and consequently
\begin{equation}
\label{eqn:Cauchy-Davenport-Generalization}
|A_1 + A_2 + \cdots + A_k| \geq \min_{d: d > 0, d | n}d\left(\left\lceil\frac{|A_1|}{d}\right\rceil + \left\lceil\frac{|A_2|}{d}\right\rceil + \cdots + \left\lceil\frac{|A_k|}{d}\right\rceil - (k-1)\right)
\end{equation}
\begin{proof}
See Appendix\ref{Appendix_Cauchy-Davenport}.
\end{proof}
\end{theorem}

It can be checked that when $k = 2$ and $n$ is a prime, equation (\ref{eqn:Cauchy-Davenport-Generalization}) in Theorem \ref{thm:Cauchy-Davenport-Generalization} degenerates to the Cauchy-Davenport Theorem. Based on Theorem \ref{thm:Cauchy-Davenport-Generalization}, we are able to further characterize the linear solvability of the network $\mathcal{N}_{\omega, \mathbf{d}}$ from the perspective of subgroup orders.

\begin{theorem}
\label{thm:Concise_Linear_Solvability_Characterization}
Consider a network $\mathcal{N}_{\omega,\mathbf{d}}$ constructed by Algorithm \ref{Alg:General_Construction} with parameters $\omega$ and $\mathbf{d} = \{d_1, d_2, \cdots, d_\omega\}$. It is linearly solvable over GF($q$) if and only if there is positive divisor $d$ of $q - 1$ subject to
\begin{equation}
\label{eqn:equivalence_subgroup_order_perspective}
q \geq d\left(\left\lceil\frac{d_1}{d}\right\rceil + \cdots + \left\lceil\frac{d_\omega}{d}\right\rceil - \omega + 1 \right) + 2
\end{equation}
or equivalently,
\begin{equation}
\label{eqn:equivalence_coset_perspective}
\frac{q-1}{d} > \left\lceil\frac{d_1}{d}\right\rceil + \cdots + \left\lceil\frac{d_\omega}{d}\right\rceil - \omega + 1
\end{equation}
\begin{proof}
As a consequence of Proposition \ref{prop:linear_solvability_exhaust_claim}, it is equivalent to show that there exist subsets $T_1, T_2, \cdots, T_\omega$ of $\mathbb{Z}_{q-1}$ with respective cardinalities $d_1, d_2, \cdots , d_\omega$ such that the set $T_1 + T_2 + \cdots + T_\omega$ does not exhausting $\mathbb{Z}_{q-1}$ if and only if there is a divisor $d > 0$ of $q - 1$ subject to (\ref{eqn:equivalence_subgroup_order_perspective}).

For the necessity part, let $T_1, T_2, \cdots, T_\omega$ be subsets of $\mathbb{Z}_{q-1}$ with respective cardinalities $d_1, d_2, \cdots , d_\omega$ subject to $|T_1 + T_2 + \cdots + T_\omega| < q-1$. According to Theorem \ref{thm:Cauchy-Davenport-Generalization},
\[
\min_{d: d > 0, d | q-1}d\left(\left\lceil\frac{|T_1|}{d}\right\rceil + \left\lceil\frac{|T_2|}{d}\right\rceil + \cdots + \left\lceil\frac{|T_\omega|}{d}\right\rceil - (\omega-1)\right) + 1 \leq |T_1 + T_2 + \cdots + T_\omega| < q-1.
\]
This implies that there exists at least one divisor $d > 0$ of $q - 1$ to make inequality (\ref{eqn:equivalence_subgroup_order_perspective}) hold.

For the sufficiency part, let $d > 0$ be an arbitrary divisor of $q-1$ such that inequality (\ref{eqn:equivalence_subgroup_order_perspective}) holds. Thus, there is a subgroup of $\mathbb{Z}_{q-1}$, to be denoted by $H$, of order $d$. For all $1 \leq j \leq \omega$, write $d_j' = \left\lceil\frac{d_j}{d}\right\rceil$ for brevity. Note that under condition (\ref{eqn:equivalence_subgroup_order_perspective}), every $d_j'$ is not greater than $(q - 1)/d$. For each $1 \leq j \leq \omega$, let $U_j$ denote the union $\{0, 1, \cdots, d_j'-1\} + H$ of $d_j'$ cosets $\{0\}+H$, $\{1\} + H$, $\cdots$, $\{d_j'-1\}+H$ of $H$ in $\mathbb{Z}_{q-1}$. Thus,
\[
U_1 + U_2 = \{0, 1, \cdots, d_1'+d_2'-2\} + H,
\]
and recursively
\[
U_1 + U_2 + \cdots + U_\omega = \{0, 1, \cdots, d_1'+d_2'+\cdots+d_\omega'-\omega\} + H.
\]
Since $\{0, 1, \cdots, d_1'+d_2'+\cdots+d_\omega'-\omega\}$ is in at most $d_1' + \cdots + d_\omega' - (\omega - 1)$ cosets of $H$,
\begin{align*}
|U_1 + U_2 + \cdots + U_\omega| &\leq \left(d_1' + d_2' + \cdots + d_\omega' - (\omega - 1)\right)d \\
& = \left(\left\lceil\frac{d_1}{d}\right\rceil + \left\lceil\frac{d_2}{d}\right\rceil + \cdots + \left\lceil\frac{d_\omega}{d}\right\rceil - (\omega - 1)\right)d < q - 1,
\end{align*}
where the last inequality holds under condition (\ref{eqn:equivalence_subgroup_order_perspective}).
If for each $1 \leq j \leq \omega$, $T_j$ is assigned to contain arbitrary $d_j$ elements in $U_j$, then
\[
|T_1 + T_2 + \cdots + T_\omega| \leq |U_1 + U_2 + \cdots + U_\omega| < q - 1.
\]
Such a selection of $T_1, T_2, \cdots, T_\omega$ makes $T_1 + T_2 + \cdots + T_\omega$ not exhaust $\mathbb{Z}_{q-1}$ as desired, so inequality (\ref{eqn:equivalence_subgroup_order_perspective}) holds. Inequality (\ref{eqn:equivalence_coset_perspective}) is a simple variation of (\ref{eqn:equivalence_subgroup_order_perspective}).
\end{proof}
\end{theorem}

In view of the explicit characterization (\ref{eqn:equivalence_coset_perspective}) in Theorem \ref{thm:Concise_Linear_Solvability_Characterization}, the more important algebraic identity that affects the linear solvability of $\mathcal{N}_{\omega, \mathbf{d}}$ is the multiplicative subgroup order in GF($q$) rather than the field size. Let $d$ be a divisor of $q-1$. Then there is a subgroup in GF($q$)$^\times$ of order $d$. The value $\frac{q-1}{d}$ in (\ref{eqn:equivalence_coset_perspective}) represents the total number of cosets of $G$ in GF($q$)$^\times$. For each $1 \leq j \leq \omega$, $d_j$ represents the out-degree of layer-3 node $v_j$ in $\mathcal{N}_{\omega,\mathbf{d}}$, and $\lceil\frac{d_j}{d}\rceil$ in (\ref{eqn:equivalence_coset_perspective}) represents the minimum number of cosets of $G$ in GF($q$)$^\times$ such that each outgoing edge of $v_j$ can be assigned a different value in these cosets. Thus, $\mathcal{N}_{\omega,\mathbf{d}}$ is the first class of multicast networks ever discovered which has a linear solvability characterization that matches the topological parameters with multiplicative subgroups and the associated coset numbers in GF($q$).

Based on Theorem \ref{thm:Concise_Linear_Solvability_Characterization}, the fact that the network $\mathcal{N}_{3,\mathbf{d}}$ with $\mathbf{d} = (5, 5, 10)$ as depicted in Fig. \ref{Fig:Rank_3_networks}(b) is not linearly solvable over GF(17) can be easily verified without exhaustive enumeration.

\vspace{5pt}

\section{A Subgroup Order Criterion}
\label{Sec:Subgroup_Order_Criterion}
One application of the explicit linear solvability characterization of the network $\mathcal{N}_{\omega,\mathbf{d}}$ constructed by Algorithm \ref{Alg:General_Construction} is to systematically yield multicast network linearly solvable over GF($q$) but not over GF($q'$) with $q < q'$, and hence $q_{min} \leq q < q' \leq q^*_{max}$. We next introduce a simple way to do so stemming from the following criterion on subgroup orders.

\begin{definition}\label{Def:Subgroup_Order_Criterion}
A pair (GF($q$), GF($q'$)) of finite fields is said to satisfy the \emph{subgroup order criterion} if there is such a proper subgroup $G$ of GF$(q)^\times$ other than $\{1\}$ that

\vspace{3pt}

\begin{Bullet}
for every proper subgroup $G'$ of GF$(q')^\times$, at least $|G| > |G'|$ or $q - |G| > q' - |G'|$.\footnote{Note that when $q > q'$, it is possible for $|G| > |G'|$ and $q - |G| > q' - |G'|$ to hold at the same time. But when $q < q'$, only one of $|G| > |G'|$ and $q - |G| > q' - |G'|$ can hold. }
\end{Bullet}
\vspace{3pt}
The inequality $q - |G| > q' - |G'|$ in $(*)$ implies the smaller cardinality of $\mathrm{GF}(q')^\times\backslash G'$ than the one of $\mathrm{GF}(q)^\times\backslash G$.
\end{definition}

It can be verified that for all known multicast networks with $q^*_{max} > q_{min}$, the corresponding (GF($q_{min}$), GF($q^*_{max}$)) satisfies the subgroup order criterion. In particular, (GF(7), GF(8)) and (GF(16), GF(17)) can be shown to satisfy the criterion by respectively setting $G$ to be any proper subgroup of GF($7$)$^\times$ other than $\{1\}$ and to be the subgroup of GF($16$)$^\times$ of order $5$. Some simple sufficient conditions for the subgroup order criterion to hold are characterized below.

\begin{proposition}
\label{Prop:Subgroup_Order_Criterion}
Consider a pair (GF($q$), GF($q'$)) with $q-1$ being a composite. If $q' < q$ or $q'$ is equal to a prime plus 1, then by setting $G$ to be an arbitrary proper subgroup of GF($q$)$^\times$ other than $\{1\}$, condition $(*)$ holds. Thus, in both cases, the pair (GF($q$), GF($q'$)) satisfies the subgroup order criterion.
\begin{proof}
When $q-1$ is a composite, GF($q$)$^\times$ has a proper subgroup other than $\{1\}$. For the case $q' < q$, let $G'$ be any proper subgroup of GF($q'$)$^\times$. If $|G'| < |G|$, then there is nothing to prove. If $|G'| \geq |G|$, then $q - |G| > q' - |G| \geq q' - |G'|$. Condition $(*)$ thus holds. %
For the case that $q'$ is a prime plus one, $\{1\}$ is the only proper subgroup of GF($q'$)$^\times$. Therefore, by setting $G$ to be any proper subgroup of GF($q$)$^\times$ other than $\{1\}$, condition ($*$) naturally holds.
\end{proof}
\end{proposition}

Based on the concept of subgroup order criterion on $(\mathrm{GF}(q), \mathrm{GF}(q'))$, parameters $\omega, \mathbf{d}$ for the general network $\mathcal{N}_{\omega,\mathbf{d}}$ can be appropriately chosen such that it has a linear solution over GF($q$) but no linear solution over GF($q'$).

\begin{theorem}
\label{Thm:Implication_Subgroup_Order_Criterion}
Consider a pair (GF($q$), GF($q'$)) subject to the subgroup order criterion. Let $G$ be any proper subgroup of GF($q$)$^\times$ satisfying $(*)$ other than $\{1\}$. Write $d = |G|$. For the multicast network $\mathcal{N}_{\omega, \mathbf{d}}$ constructed by Algorithm \ref{Alg:General_Construction}, select $\omega \geq 3$ to satisfy
\begin{equation}
\label{eqn:omega_sharp_bound}
\omega \geq \max\left\{\frac{\frac{q' - 1}{d'} - \
\lceil\frac{q-d-1}{d'}\rceil}{\lceil\frac{d}{d'}\rceil-1} + 1: 1 \leq d' < d, d' | q' -1\right\}
\end{equation}
and set the $\omega$-tuple $\mathbf{d}$ to be $(\underbrace{d, d, \cdots, d}_{\omega-1}, q-d-1)$. Then, $\mathcal{N}_{\omega, \mathbf{d}}$ is linearly solvable over GF($q$) but not over GF($q'$). Moreover, $q_{min} = q$ for the considered $\mathcal{N}_{\omega,\mathbf{d}}$.
\begin{proof}
First recall that Algorithm \ref{Alg:General_Construction} requires $\omega \geq 3$ for the construction of $\mathcal{N}_{\omega,\mathbf{d}}$. Since $d$ is the order of subgroup $G$ of GF($q$)$^\times$, $d$ divides $q - 1$ as well as $q-d-1$. Thus,
\begin{align*}
&~~~~d\left(\underbrace{\left\lceil\frac{d}{d}\right\rceil + \cdots + \left\lceil\frac{d}{d}\right\rceil}_{\omega-1} + \left\lceil\frac{q - d -1}{d}\right\rceil - \omega + 1 \right) + 2 \\
&= d\cdot \frac{q - d -1}{d} + 2 \\
&= q - d + 1 \leq q,
\end{align*}
and condition (\ref{eqn:equivalence_subgroup_order_perspective}) in Theorem \ref{thm:Concise_Linear_Solvability_Characterization} holds. The considered network is thus linearly solvable over GF($q$).

Let $G'$ be an arbitrary subgroup of GF($q'$)$^\times$. Write $d' = G'$.

In the case $d > d'$,
\begin{align*}
&~~~~d'\left(\underbrace{\left\lceil\frac{d}{d'}\right\rceil + \cdots + \left\lceil\frac{d}{d'}\right\rceil}_{\omega-1} + \left\lceil\frac{q - d - 1}{d'}\right\rceil - \omega + 1 \right) + 2 \\
&= d' \left((\omega-1)\left(\left\lceil\frac{d}{d'}\right\rceil - 1\right) + \left\lceil\frac{q-d-1}{d'}\right\rceil\right) + 2 \\
&\geq d' \left(\frac{q'-1}{d'} - \left\lceil\frac{q - d - 1}{d'}\right\rceil + \left\lceil\frac{q-d-1}{d'}\right\rceil \right) + 2 \\
&= q' + 1,
\end{align*}
where the inequality holds due to the assumed value of $\omega$. Thus, condition (\ref{eqn:equivalence_subgroup_order_perspective}) does not hold over GF($q'$) for $d'$ in this case.

In the case $d \leq d'$ but $q - d > q' - d'$,
\begin{align*}
&~~~~d'\left(\underbrace{\left\lceil\frac{d}{d'}\right\rceil + \cdots + \left\lceil\frac{d}{d'}\right\rceil}_{\omega-1} + \left\lceil\frac{q - d - 1}{d'}\right\rceil - \omega + 1 \right) + 2 \\
& = d'\left\lceil\frac{q - d - 1}{d'}\right\rceil + 2 \\
&\geq d' \left(\frac{q' - d' - 1}{d'} + 1\right) + 2 = q' + 1,
\end{align*}
where the last inequality can be established by noting that $q - d > q' - d'$ and $d'$ divides $q' - d' -1$. Thus, condition  (\ref{eqn:equivalence_subgroup_order_perspective}) does not hold over GF($q'$) for $d'$ in this case either.

It has been verified that condition (\ref{eqn:equivalence_subgroup_order_perspective}) does not hold over GF($q'$) for any divisor $d' > 0$ of $q' - 1$. Theorem \ref{thm:Concise_Linear_Solvability_Characterization} then asserts the considered network $\mathcal{N}_{\omega,\mathbf{d}}$ is not linearly solvable over GF($q'$).

Now consider an arbitrary prime power $q'' < q$ and an arbitrary divisor $d'' > 0$ of $q''-1$. In the case $q - d > q'' - d''$, similar to the analysis above,
\[
d''\left(\underbrace{\left\lceil\frac{d}{d''}\right\rceil + \cdots + \left\lceil\frac{d}{d''}\right\rceil}_{\omega-1} + \left\lceil\frac{q - d - 1}{d''}\right\rceil - \omega + 1 \right) + 2 \geq d''\left\lceil\frac{q - d - 1}{d''}\right\rceil + 2 \geq q'' + 1,
 \]
so condition (\ref{eqn:equivalence_subgroup_order_perspective}) does not hold over GF($q''$) for $d''$. Consider the case $q - d \leq q'' - d''$. This implies $d > d''$ and hence $\left\lceil\frac{d}{d''}\right\rceil \geq 2$. Note that
\[
\left\lceil\frac{q-d-1}{d''}\right\rceil \geq \left\lceil\frac{q''-d-1}{d''}\right\rceil \geq \frac{q'' - 1}{d''} - \left\lceil\frac{d}{d''}\right\rceil.
\]
Consequently,
\begin{align*}
&~~~~d''\left(\underbrace{\left\lceil\frac{d}{d''}\right\rceil + \cdots + \left\lceil\frac{d}{d''}\right\rceil}_{\omega-1} + \left\lceil\frac{q - d - 1}{d''}\right\rceil - \omega + 1 \right) + 2 \\
&\geq d''\left((\omega-1)\left(\left\lceil\frac{d}{d''}\right\rceil - 1 \right) + \frac{q'' - 1}{d''} - \left\lceil\frac{d}{d''}\right\rceil \right) + 2\\
&\geq q'' + 1,
\end{align*}
where the last inequality holds due to $\omega \geq 3$ and $\left\lceil\frac{d}{d''}\right\rceil \geq 2$. Hence, condition (\ref{eqn:equivalence_subgroup_order_perspective}) does not hold over GF($q''$) for $d''$ in this case either. We conclude that the considered network is not linearly solvable over GF($q''$) and hence $q_{min} = q$. \end{proof}
\end{theorem}

\vspace{5 pt}

\begin{remark}
In the theorem above, the bound (\ref{eqn:omega_sharp_bound}) on the choice of $\omega$ to guarantee the network $\mathcal{N}_{\omega,\mathbf{d}}$ not linearly solvable over GF($q'$) is sharp, since if $\omega < \frac{\frac{q' - 1}{d'} - \
\lceil\frac{q-d-1}{d'}\rceil}{\lceil\frac{d}{d'}\rceil-1} + 1$ for some divisor $d'$ of $q' - 1$ smaller than $d$, then
\begin{align*}
&~~~~d'\left(\underbrace{\left\lceil\frac{d}{d'}\right\rceil + \cdots + \left\lceil\frac{d}{d'}\right\rceil}_{\omega-1} + \left\lceil\frac{q - d - 1}{d'}\right\rceil - \omega + 1 \right) + 2 \\
& = d'\left((\omega-1)\left(\left\lceil\frac{d}{d'}\right\rceil - 1\right) + \left\lceil\frac{q - d - 1}{d'}\right\rceil \right)  + 2 \\
& < q' + 1,
\end{align*}
and thus Theorem \ref{thm:Concise_Linear_Solvability_Characterization} affirms that the network is linearly solvable over GF($q'$).

Besides the sharp bound (\ref{eqn:omega_sharp_bound}) on $\omega$, another convenient bound of $\omega ( \geq 3)$ to guarantee the network $\mathcal{N}_{\omega,\mathbf{d}}$ not linearly solvable over GF($q'$) is
\begin{equation}
\label{eqn:omega_weak_bound}
\omega \geq \frac{q'-q+d}{d-d'_{max}}+1,
\end{equation}
where $d'_{max}$ is the largest divisor of $q' - 1$ that is smaller than $d$. Condition (\ref{eqn:omega_weak_bound}) implies (\ref{eqn:omega_sharp_bound}) due to $\frac{\frac{q' - 1}{d'} - \
\lceil\frac{q-d-1}{d'}\rceil}{\lceil\frac{d}{d'}\rceil-1} < \frac{\frac{q' - 1}{d'} - \frac{q-d-1}{d'}}{\frac{d}{d'}-1} = \frac{q' - q + d}{d - d'}$
for any $d' < d$. When $q' - 1$ is a prime, that is, when $1$ is the only proper divisor $q' - 1$, conditions (\ref{eqn:omega_sharp_bound}) and (\ref{eqn:omega_weak_bound}) become the same. In general, rule (\ref{eqn:omega_weak_bound}) is not tight. For example, consider the pair (GF(16), GF(17)) which satisfies the subgroup order criterion by setting $G$ to be the subgroup of GF(16)$^\times$ containing 5 elements. Thus, the smallest $\omega$ subject to (\ref{eqn:omega_sharp_bound}) is 2, but (\ref{eqn:omega_weak_bound}) requires $\omega \geq 7$. \hfill $\blacksquare$
\end{remark}

\begin{figure}[htbp]
\centering
\scalebox{0.5}
{\includegraphics{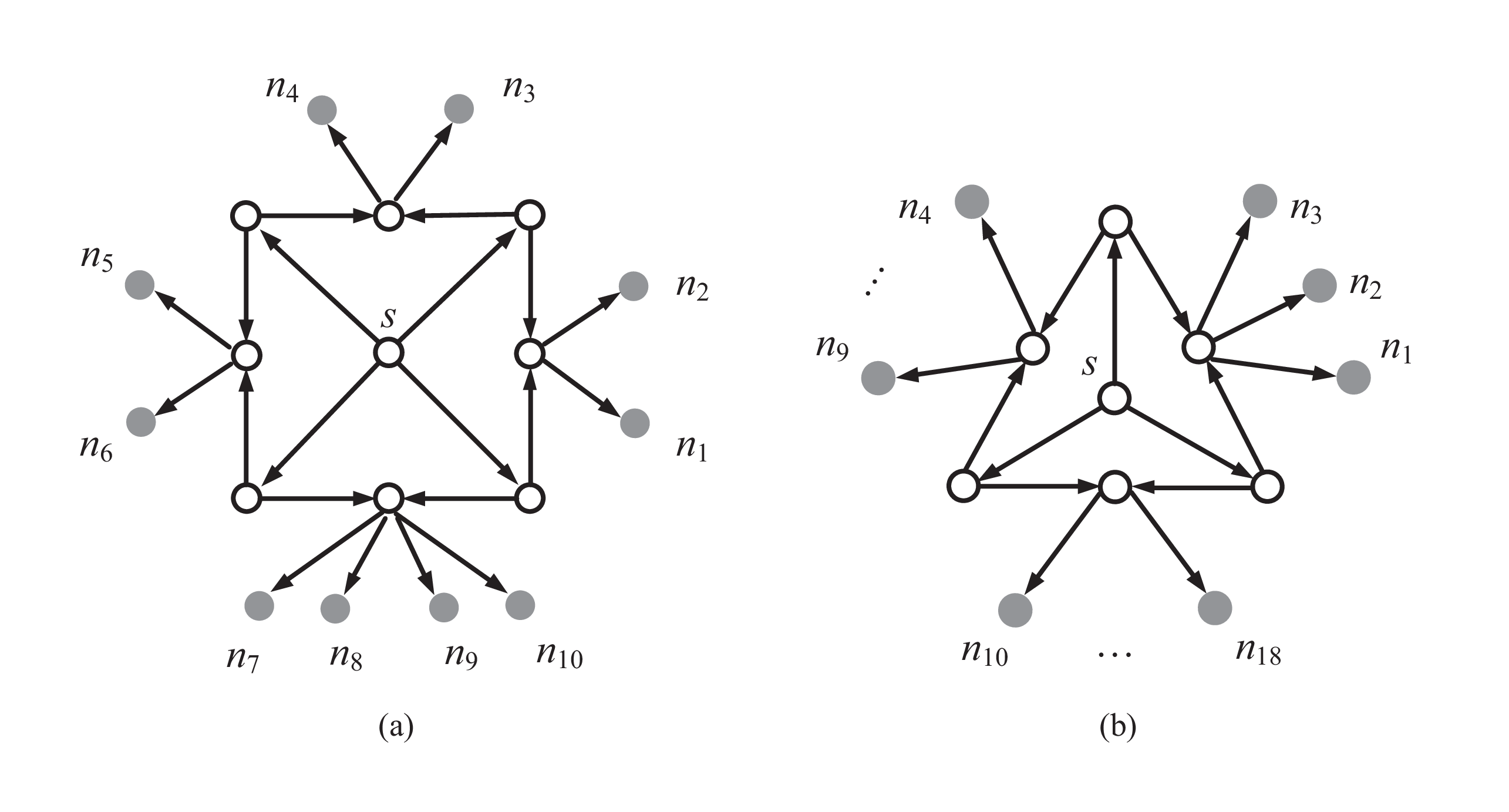}}
\caption{The networks $\mathcal{N}_{\omega, \mathbf{d}}$ constructed by Algorithm \ref{Alg:General_Construction} have respective parameters (a) $\omega = 4$, $\mathbf{d} = (2, 2, 2, 4)$; (b) $\omega = 3$, $\mathbf{d} = (3, 6, 9)$. The network in (a) is a different one linearly solvable over GF(7) but not over GF(8) from the one depicted in Fig. \ref{Fig:Rank_3_networks}(a). The network in (b) is a different one linearly solvable over GF(16) but not over GF(17) from the one depicted in Fig. \ref{Fig:Rank_3_networks}(b).}
\label{Fig:Rank_4_network}
\end{figure}

For a same pair (GF($q$), GF($q'$)) satisfying the subgroup order criterion, different choices of subgroups $G$ of GF($q$)$^\times$ obeying $(*)$ will yield different multicast networks $\mathcal{N}_{\omega,\mathbf{d}}$ linearly solvable over GF($q$) but not over GF($q'$). For instance, for the pair (GF(7), GF(8)), both subgroups $G = \{1, 2, 4\} \subset \mathrm{GF}(7)^\times$ and $G = \{1, 6\} \subset \mathrm{GF}(7)^\times$ satisfy condition $(*)$. Hence, according to Theorem \ref{Thm:Implication_Subgroup_Order_Criterion}, not only the network $\mathcal{N}_{3, \mathbf{d}}$ with $\mathbf{d} = (3, 3, 3)$ depicted in Fig. \ref{Fig:Rank_3_networks}(a), but also the network $\mathcal{N}_{4, \mathbf{d}}$ with $\mathbf{d} = (2, 2, 2, 4)$ depicted in Fig. \ref{Fig:Rank_4_network}(a) can be constructed to be linearly solvable over GF(7) but not over GF(8).

Based on the subgroup order criterion and Theorem \ref{Thm:Implication_Subgroup_Order_Criterion}, a number of interesting new multicast networks $\mathcal{N}_{\omega,\mathbf{d}}$ linearly solvable over GF($q$) but not over GF($q'$) can be accordingly designed. This will be illustrated in the next section. However, it is worth noting that \emph{not} every $\mathcal{N}_{\omega,\mathbf{d}}$ with $q_{min} < q^*_{max}$ can be found in this manner.
For instance, consider the pair (GF(16), GF(17)). Though it satisfies the subgroup order criterion with respect to $G$ being the subgroup of order 5 in GF($16$)$^\times$, if we set $G$ to be the subgroup of order 3 in GF($16$)$^\times$, condition $(*)$ is no longer obeyed. On the other hand, one can check, based on Theorem \ref{thm:Concise_Linear_Solvability_Characterization}, that the network $\mathcal{N}_{3, \mathbf{d}}$ depicted in Fig. \ref{Fig:Rank_4_network}(b) with $\mathbf{d} = (3, 6, 9)$ is also linearly solvable over GF(16) but not over GF(17). Stemming from Theorem \ref{thm:Concise_Linear_Solvability_Characterization}, one might be able to formulate other criteria from the perspective of subgroup orders for appropriate selection of $\omega$ and $\omega$-tuple $\mathbf{d}$ such that the corresponding network $\mathcal{N}_{\omega,\mathbf{d}}$ has $q_{min} < q^*_{max}$. However, that is beyond the scope of the current paper, since as we shall see in the next section, the subgroup order criterion discussed in this section has already allowed us to unveil infinitely many new interesting instances with $q_{min} < q^*_{max}$.

\vspace{5pt}

\section{Construction of Multicast Networks with $q_{min} < q^*_{max}$}
\label{Sec:Establish_Instances}
In the following, we aim at finding new multicast networks with $q_{min} < q^*_{max}$ and answering open problems raised in \cite{Sun_TIT}. The tool is to establish prime power pairs (GF($q$), GF($q'$)) with $q < q'$ subject to the subgroup order criterion introduced in the previous section.

As pointed out in \cite{Sun_TIT}, one of their motivations to design the first few exemplifying multicast networks with $q_{min} < q^*_{max}$ is the matroid structure of \emph{free swirl}. Correspondingly, a class of so-called \emph{Swirl networks} are constructed, which are linearly solvable over GF($q_{min}$) with $q_{min} = 5$ but not over GF($q'$) when $q' \leq \omega+2$ and $q'$ is equal to a prime plus 1. As special instances in the present framework, they are the networks $\mathcal{N}_{\omega,\mathbf{d}}$ constructed by Algorithm \ref{Alg:General_Construction} with an arbitrary source dimension $\omega \geq 3$ and the $\omega$-tuple $\mathbf{d} = (2, 2, \cdots, 2)$. When the prime power $q' > 3$ is equal to a prime plus 1, it must be in the form of $2^p$ with $p$ a prime and $2^p - 1$ is known to be a Mersenne prime. Since there exist $q_1$ and $q_2$ such that $2^{q_1} - 1$ and $2^{q_2} - 1$ are composite while $2^{q_1 + q_2} - 1$ is a Mersenne prime, the Swirl network not only demonstrates that $q_{min} < q^*_{max}$ is possible for multicast networks, but also verifies a conjecture raised in \cite{Fragouli11} that a multicast network linearly solvable over GF($2^{q_1}$) and GF($2^{q_2}$) is \emph{not even necessarily} linearly solvable over GF($2^{q_1+q_2}$). However, although it is a popular conjecture that there are infinitely many Mersenne primes in the literature, only 48 Mersenne primes have been discovered so far (See the GIMPS project \cite{GIMPS}). Thus, at this moment, the class of Swirl networks, as well as other exemplifying networks with $q_{min} < q^*_{max}$, still fails to show the following fundamental problems on multicast networks:
\begin{itemize}
\item Can GF($q_{min}$), GF($q^*_{max}$) have the \emph{same} characteristic?
\item Can the gap $q^*_{max} - q_{min} > 0$ tend to \emph{infinity}?
\item Are there \emph{infinitely many} prime power pairs $(q, q')$ with $q < q'$ such that each $(q, q')$ is equal to $(q_{min}, q^*_{max})$ for some multicast networks?
\end{itemize}

In the remainder of this section, we first design appropriate parameters $\omega$ and $\mathbf{d}$ for the general network $\mathcal{N}_{\omega,\mathbf{d}}$ to affirm positive answers to \emph{all} problems listed above.

\begin{theorem}
\label{thm:multicast_networks_characteristic_2}
Let $q = 2^{2k}$ and $d = \frac{2^{2k}-1}{3}$, where $k$ is an arbitrary integer larger than 2. Set $\omega \geq 3$ satisfying (\ref{eqn:omega_sharp_bound}) and the $\omega$-tuple $\mathbf{d} = (\underbrace{d, \cdots, d}_{\omega-1}, 2d)$. Then, the multicast network $\mathcal{N}_{\omega,\mathbf{d}}$ is linearly solvable over GF($q$) but \emph{not} over GF($2q$). Moreover, $q_{min} = 2^{2k}$ and $q^*_{max} = 2^{2k'+1}$ for some $k' \geq k$.
\begin{proof}
We shall first show that the subgroup order criterion of Definition \ref{Def:Subgroup_Order_Criterion} holds for (GF($q$), GF($2q$)). Since $2^{2k}-1$ is divisible by 3, in the multiplicative group $\mathrm{GF}(2^{2k})^\times$, there is a subgroup $G$ of order $d = \frac{2^{2k} - 1}{3}$. On the other hand, it is easy to check that 7 is the smallest integer larger than 1 that divides $2^{2k+1} - 1$. Let $d'$ be the order of an arbitrary subgroup $G'$ of $\mathrm{GF}(2^{2k+1})^\times$. Then, $d' \leq \frac{2^{2k+1} - 1}{7}$. Because
\[
d = \frac{2^{2k}-1}{3} = \frac{2^{2k+1} - 2}{6} > \frac{2^{2k+1}-1}{7} \geq d',
\]
we conclude that condition $(*)$ in Definition \ref{Def:Subgroup_Order_Criterion} holds and the subgroup order criterion is satisfied. Moreover, since $\omega$ is set to satisfy (\ref{eqn:omega_sharp_bound}) and $2d = 2^{2k-1} - d - 1$, Theorem \ref{Thm:Implication_Subgroup_Order_Criterion} asserts the considered network is linearly solvable over GF($2^{2k}$) but not over GF($2^{2k+1}$) and $q_{min} = 2^{2k}$.

It remains to prove $q^*_{max} = 2^{2k'+1}$ for some $k' \geq k$. Let $q'$ be an arbitrary prime power larger than $2^{2k+1}$ and \emph{not} in the form of $2^{2k'+1}$. It is then equivalent to show that the network is linearly solvable over GF($q'$). As a consequence of the equivalent condition (\ref{eqn:equivalence_subgroup_order_perspective}) in Theorem \ref{Thm:Implication_Subgroup_Order_Criterion}, it suffices to find a proper subgroup $G'$ of $\mathrm{GF}(q')^\times$ such that
\begin{equation}
\label{eqn:Prove_q*_max}
|G'| \geq d~\mathrm{and}~q'-|G'| \geq q - d,
\end{equation}
because under such a choice, condition (\ref{eqn:equivalence_subgroup_order_perspective}) holds as
\begin{align*}
&~~~|G'|\left(\underbrace{\left\lceil\frac{d}{|G'|} \right\rceil + \cdots + \left\lceil\frac{d}{|G'|} \right\rceil}_{\omega-1} + \left\lceil\frac{q-d-1}{|G'|}\right\rceil - \omega + 1 \right) + 2 \\
&< |G'| + q - d + 1 \leq q' + 1
\end{align*}
and Theorem \ref{thm:Concise_Linear_Solvability_Characterization} then implies the considered network is linearly solvable over GF($q'$).

Assume that $q'$ is odd. Set $G'$ to be the subgroup $\{\xi^2, \xi^4, \cdots, \xi^{q'-1}\}$ of $\mathrm{GF}(q')^\times$, where $\xi$ is a primitive element of GF($q'$). It can be checked that
\[
|G'| = \frac{q'-1}{2} \geq 2^{2k} > d, q' - |G'| > 2^{2k} > q - d - 1,
\]
and hence condition (\ref{eqn:Prove_q*_max}) holds.

Assume that $q'$ is even. Since $q'$ is assumed not in the form of $2^{2k'+1}$, $q' = 2^{2k'}$ for some $k' > k$. Then, $q' - 1$ is divisible by 3 and hence we can set $G'$ to be the subgroup $\{\xi^3, \xi^6, \cdots, \xi^{q'-1}\}$ of $\mathrm{GF}(q')^\times$. It can be checked
\begin{eqnarray*}
|G'| &=& \frac{q'-1}{3} > \frac{q-1}{3} = d, \\
q' - |G'| - 1 &=& \frac{2}{3}(q'-1) > \frac{2}{3}(q-1) = q - d - 1,
\end{eqnarray*}
so condition (\ref{eqn:Prove_q*_max}) holds for this case too. We have verified that $q^*_{max}$ must be in the form $2^{2k'+ 1}$ form some $k' \geq k$.
\end{proof}
\end{theorem}

The class of networks $\mathcal{N}_{\omega,\mathbf{d}}$ under the setting of $\omega$ and $\mathbf{d}$ in Theorem \ref{thm:multicast_networks_characteristic_2} not only turns out to be the first discovered in the literature with GF($q_{min}$), GF($q^*_{max}$) having the \emph{same characteristic}, but also subsequently answers two open questions raised in \cite{Sun_TIT}.

\begin{corollary}
\label{coro:infi_even_prime_powers}
There are \emph{infinitely many even} prime power pairs $(q, q')$ such that $q = q_{min} < q^*_{max} = q'$ for some multicast network. Moreover, the gap $q^*_{max} - q_{min} > 0$ can tend to \emph{infinity}.
\begin{proof}
Consider the network $\mathcal{N}_{\omega,\mathbf{d}}$ under the setting of $\omega$ and $\mathbf{d}$ in Theorem \ref{thm:multicast_networks_characteristic_2}. When $k$ tends to infinity, $q^*_{max} - q_{min} \geq 2^{2k+1} - 2^{2k} = 2^{2k}$ tends to infinity too.
\end{proof}
\end{corollary}

Theorem \ref{thm:multicast_networks_characteristic_2} uncovered infinitely many pairs (GF($q$), GF($q'$)) of finite fields with characteristic $2$ and $q < q'$ such that there is a multicast network $\mathcal{N}_{\omega,\mathbf{d}}$ linearly solvable over GF($q$) but not over GF($q'$). We next further extend the result to the cases that GF($q$) and GF($q'$) have \emph{arbitrary distinct} characteristics. We need the following lemma, which is established based on the \emph{Weyl's equidistribution theorem} (See, e.g., \cite{Book_for_Equidistribution}.)

\begin{lemma}
\label{lemma:Diophantine_approximation}
Let $n_1, n_2$ be two coprime integers larger than 1, and $c_1$, $c_2$, $\delta$ be real numbers with $c_1 > c_2$. There are infinitely many positive integers $k, k'$ such that
\begin{equation}
\label{eqn:Diophantine_approximation}
\log_{n_1}n_2^{k'} + c_1 > k > \log_{n_1}(n_2^{k'} + \delta) + c_2
\end{equation}
\begin{proof}
See Appendix\ref{Appendix_Diophantine}.
\end{proof}
\end{lemma}

\begin{theorem}
\label{thm:p_vs_p'_general_odd}
Let $p, p'$ be two arbitrary distinct prime numbers. There are \emph{infinitely many} $k, k'$ such that there is a multicast network $\mathcal{N}_{\omega,\mathbf{d}}$ linearly solvable over GF($q_{min}$) with $q_{min} = p^{k}$, but \emph{not} over GF($p'^{k'}$) with $p^{k} < p'^{k'}$.
\begin{proof}
(Sketch) Note that there must be such a prime power $q = p^j$ that $q - 1$ has a divisor $d$ no smaller than 3. Then, $q^k - 1$ is divisible by $d$ for all $k \geq 1$. Based on Lemma \ref{lemma:Diophantine_approximation}, it can be shown that there are infinitely many integers $k, k'$ with $q^k < p'^{k'}$ such that the pair (GF($q^k$), GF($q'^{k'}$)) satisfies the subgroup order criterion by setting $G$ to be the subgroup of GF($q^k$)$^\times$ of order $\frac{q^k-1}{d}$, and hence there is a multicast network $\mathcal{N}_{\omega,\mathbf{d}}$ linearly solvable over GF($q_{min}$) with $q_{min} = q^{k}$, but \emph{not} over GF($q'^{k'}$) according to Theorem \ref{Thm:Implication_Subgroup_Order_Criterion}. %
A detailed proof can be found in Appendix\ref{Appendix_General_p_vs_p}.
\end{proof}
\end{theorem}

As a counterpart of Theorem \ref{thm:multicast_networks_characteristic_2}, Theorem \ref{thm:p_vs_p'_general_odd} unveils that for \emph{any} two distinct primes $p$ and $p'$, we can make use of the subgroup order criterion to find infinitely many ($p^k$, $p'^{k'}$) with $p^k < p'^{k'}$ such that there is a multicast network $\mathcal{N}_{\omega,\mathbf{d}}$ linearly solvable over GF($p^k$) but not over GF($p'^{k'}$). However, $q^*_{max}$ is not necessarily a power of $p'$ for this established $\mathcal{N}_{\omega,\mathbf{d}}$, which is weaker than the consideration in Theorem \ref{thm:multicast_networks_characteristic_2}. However, we can still have the following partial generalization of Corollary \ref{coro:infi_even_prime_powers}, as a consequence of the above theorem.

\begin{corollary}
\label{corollary:infi_general_prime_powers}
Let $p$ be an arbitrary prime. There are \emph{infinitely many} prime power pairs $(p^k, q')$ such that $q_{min} = p^k < q' = q^*_{max}$ for some multicast networks.
\end{corollary}

All above interesting results are established based on the general framework developed in Section \ref{Sec:General_Framework} and the subgroup order criterion formulated in Section \ref{Sec:Subgroup_Order_Criterion}. Yet, this approach seems not helpful to find exemplifying multicast networks linearly solvable over GF($q$) but not over GF($q'$) with $q < q'$ when GF($q$) and GF($q'$) have the \emph{same odd} characteristic. A key reason is as follows.

\begin{proposition}
\label{prop:framework_insufficiency}
If the network $\mathcal{N}_{\omega,\mathbf{d}}$ constructed by Algorithm \ref{Alg:General_Construction} is linearly solvable over GF($p^k$), where $p$ is odd, then it is linearly solvable over GF($p^{k'}$) for all $k' \geq k$.
\begin{proof}
If the network $\mathcal{N}_{\omega,\mathbf{d}}$ is linearly solvable over GF($p^k$), then there is a divisor $d > 0$ of $p^k - 1$ subject to $p^k \geq d\left(\left\lceil\frac{d_1}{d}\right\rceil + \cdots + \left\lceil\frac{d_\omega}{d}\right\rceil - \omega + 1 \right) + 2$ according to Theorem \ref{thm:Concise_Linear_Solvability_Characterization}. This implies that $p^k \geq (d_1 - 1) + \cdots + (d_\omega - 1) + d + 2$, and thus $p^k \geq d_j - 1$ for all $1 \leq j \leq \omega$. %
Let $k'$ be an arbitrary integer larger than $k$. Since $p$ is odd, $p^{k'} - 1$ is divisible by 2. Write $d' = \frac{p^{k'} - 1}{2}$. Since $d' > p^k$, we have $d' \geq d_j$ for all $1 \leq j \leq \omega$. Hence,
\[
d'\left(\left\lceil\frac{d_1}{d'}\right\rceil + \cdots + \left\lceil\frac{d_\omega}{d'}\right\rceil - \omega + 1 \right) + 2 = d' + 2 < p^{k'}.
\]
Theorem \ref{thm:Concise_Linear_Solvability_Characterization} now in turn affirms that the network $\mathcal{N}_{\omega,\mathbf{d}}$ is linearly solvable over GF($p^{k'}$).
\end{proof}
\end{proposition}

\vspace{3pt}

\noindent\textbf{Problem}. Is there a multicast network that is linearly solvable over GF($p^k$) but not over GF($p^{k'}$) for an odd prime $p$ and $k < k'$?

\vspace{5pt}

\section{Construction of Multicast Networks with Prescribed $q_{min}$}
\label{Sec:Construction_Network_Prescribed_q_min}
Let $q$ be an arbitrary prime power. The $(q+1, 2)$-combination network is the best known network in the network coding literature with $q_{min}$ equal to the prescribed $q$. Another application of the general framework developed in Section \ref{Sec:General_Framework} is to construct new multicast networks with $q_{min}$ equal to $q$.

\begin{proposition}
Assume that $q$ is not equal to a prime plus 1. Let $d$ denote an arbitrary proper divisor of $q - 1$ other than 1. The network $\mathcal{N}_{3,\mathbf{d}}$ with $\mathbf{d} = (d, d, q - d - 1)$ constructed by Algorithm \ref{Alg:General_Construction} has $q_{min} = q$.

\begin{proof}
According to Proposition \ref{Prop:Subgroup_Order_Criterion}, (GF($q$), GF($q'$)) is subject to the subgroup order criterion for all prime powers $q' < q$. Moreover, for every prime power $q' < q$, when $\omega$ is set to 3, condition (\ref{eqn:omega_sharp_bound}) in Theorem \ref{Thm:Implication_Subgroup_Order_Criterion} holds. Thus, Theorem \ref{Thm:Implication_Subgroup_Order_Criterion} shows $q_{min} = q$ for the considered network $\mathcal{N}_{3,\mathbf{d}}$ with $\mathbf{d} = (d, d, q - d - 1)$ .
\end{proof}
\end{proposition}

The proposition above justifies a new way to construct multicast networks with $q_{min}$ equal to a prescribed prime power $q$. In order to further reduce the network size while keeping $q_{min} = q$ for the constructed network, we are motivated to consider the network $\mathcal{N}_{3, \mathbf{d}}$ with $\mathbf{d} = (1, 1, q - d - 1)$, which can be regarded as a degenerate instance constructed by Algorithm \ref{Alg:General_Construction} in the sense that not every component in $\mathbf{d}$ is larger than 1. None of the previous discussions indicates the linear solvability of this degenerate case. Actually, this network has $q_{min} < q$ because in this degenerate case, condition a) becomes weaker than b) and c) in Lemma \ref{lemma:general_network_linear_solvability}.\footnote{This is the reason that the discussions in previous sections assume every component in $\mathbf{d}$ larger than 1.} We next refine the network by carefully resetting some receivers, to satisfy: (i) $q_{min} = q$, and (ii) the resulting network has a smaller number of nodes than the $(q+1, 2)$-combination network.

\begin{algorithm}
\label{alg:q-1-omega}
Let $q$ be an arbitrary prime power no smaller than 5. Construct a network, to be denoted by $\mathcal{N}_3$, consisting of nodes on $5$ layers as follows. The nodes at the first 4 layers and the edges among them are identical to the degenerate network $\mathcal{N}_{3, \mathbf{d}}$ with $\mathbf{d} = (1, 1, q - d - 1)$ constructed by Algorithm \ref{Alg:General_Construction}, and are redrawn in Fig. \ref{Fig:Netowrk_q_1_3}. Create bottom-layer nodes, which will be set as receivers, according to the following rule:
\begin{itemize}
\item There is a node connected from $\{n_{1, 1}, u_2, u_3\}$ as well as a node from $\{n_{2, 1}, u_1, u_2\}$;
\item There is a node connected from $\{n_{3, j}, u_2, u_3\}$ as well as a node from $\{n_{3, j}, u_1, u_2\}$ for each $1 \leq j \leq q-2$;
\item There is a node connected from $\{n_{1,1}, n_{3, 1}, n_{3, j}\}$ for each $1 < j \leq q-2$, and a node connected from $\{n_{2, 1}, n_{3, i}, n_{3, j}\}$ for each $1 < i < j \leq q-2$;
\item There is a node connected from $\{n_{1, 1}, n_{2,1}, n_{3, j}\}$ for each $1 \leq j \leq q - 2$.
\end{itemize}
Thus, there will be edges directly connected from layer-2 nodes to layer-5 nodes, and the total number of layer-5 nodes is $2 + (q-2) + \left(\begin{smallmatrix} q-2 \\ 2 \end{smallmatrix}\right)+ (q-2) = \frac{1}{2}q^2 - \frac{1}{2}q + 1$.
\end{algorithm}

\begin{figure}[htbp]
\centering
\scalebox{0.6}
{\includegraphics{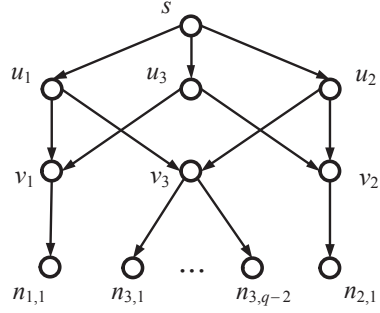}}
\caption{The first 4 layers in the network $\mathcal{N}_q$ constructed by Algorithm \ref{alg:q-1-omega}. There are additional $\frac{1}{2}q^2 - \frac{1}{2}q + 1$ non-depicted bottom-layer receivers. The multicast network has $q_{min} = q$.}
\label{Fig:Netowrk_q_1_3}
\end{figure}

\begin{proposition}
The network $\mathcal{N}_q$ constructed by Algorithm \ref{alg:q-1-omega} is linearly solvable over GF($q'$) with every $q' \geq q = q_{min}$.
\begin{proof}
In an essentially same way to derive equivalent conditions b) and c) of linear solvability for network $\mathcal{N}_{\omega,\mathbf{d}}$ in Lemma \ref{lemma:general_network_linear_solvability}, it is straightforward to show that there is a linear solution over GF($q'$) for network $\mathcal{N}_q$ if and only if there is a matrix completion of
\[
\mathbf{M}  = \left[
\begin{matrix}
1 & 0 & x_{31} &  & x_{3q} \\
x_1 & 1 & 0 & \cdots  & 0 \\
0 & x_2 & 1 &  & 1
\end{matrix}
\right]
\]
over GF($q'$), and consequently if and only if there exist $a_{1}, a_{2}, \delta_{k} \in \mathrm{GF}(q')^\times$, $1 \leq k \leq q - 2$ such that
\begin{align*}
\delta_{k} &\neq \delta_{k'}, \forall 1 \leq k < k' \leq q - 2 \nonumber \\
1 &\notin \left\{\alpha\delta_k: \alpha \in \left\{1, a_1, a_2, a_1a_2\right\}, 1 \leq k \leq q-2\right\}.
\end{align*}
When $q_0 < q$, the number of elements in $\mathrm{GF}(q_0)^\times\backslash\{1\}$ is smaller than $q - 2$, so that the condition above cannot be satisfied. When $q_0 \geq q$, set $a_1 = a_2 = 1$, and assign arbitrary $q - 2$ distinct elements in $\mathrm{GF}(q_0)^\times\backslash\{1\}$ to $\delta_1, \cdots, \delta_{q-2}$. In this way, the condition above can be satisfied. We can then conclude that the network is linearly solvable over GF($q_0$) for all $q_0 \geq q$ while not linearly solvable over GF($q_0$) for all $q_0 < q$.
\end{proof}
\end{proposition}

For the network $\mathcal{N}_q$ constructed by Algorithm \ref{alg:q-1-omega}, there are $q + 7$ nodes on the first 4 layers and then $(q + 7) + (\frac{1}{2}q^2 - \frac{1}{2}q + 1) = \frac{1}{2}q^2 + \frac{1}{2}q + 8$ nodes in the whole network $\mathcal{N}_q$. It can also be counted that there are $(q+9) + 3(\frac{1}{2}q^2 - \frac{1}{2}q + 1) = \frac{3}{2}q^2 - \frac{1}{2}q + 12$ edges in the network. In comparison, in the $(q+1, 2)$-combination network with $q_{min} = q$, there are total $2 + (q + 1) + \left(\begin{smallmatrix} q + 1 \\ 2 \end{smallmatrix}\right) = \frac{1}{2}q^2 + \frac{3}{2}q + 3$ nodes and $2 + (q + 1) + 2\left(\begin{smallmatrix} q + 1 \\ 2 \end{smallmatrix}\right) = q^2 + 2q + 3$ edges. Thus, the size of the network $\mathcal{N}_q$ is smaller than the $(q+1,2)$-combination network in terms of the number of receivers as well as the number of nodes. Although the $(q+1,2)$-combination network has a smaller number of edges, it has source dimension $\omega$ equal to 2 while the network $\mathcal{N}_q$ has $\omega = 3$. It would be fairer for the network size comparison if we increase $\omega$ of the combination network to 3. To do so, as what has been adopted in \cite{Feder03}, simply create a new node with the unique incoming edge leading from the source, and connect it to every receiver by a new edge. In this extended network, there are $\frac{1}{2}q^2 + \frac{3}{2}q + 4$ nodes and $q^2 + 2q + 4 + \left(\begin{smallmatrix} q + 1 \\ 2 \end{smallmatrix}\right) = \frac{3}{2}q^2 + \frac{5}{2}q + 4$ edges. Compared with this network, the network $\mathcal{N}_q$ constructed by Algorithm \ref{alg:q-1-omega} has not only fewer receivers and nodes, but fewer edges as well. Table \ref{table:network_size_comparison} summarizes the sizes of the discussed three types of multicast networks with $q_{min} = q$.

\begin{table}\centering
\caption{Comparison of Multicast Networks with $q_{min}$ Equal to a Prescribed Prime Power $q$}
\label{table:network_size_comparison}
\begin{tabular}{c|c|c|c}
  \hline
  \multirow{2}{*}{$~$} & $\mathcal{N}_q$ by & (q+1,2)-combination & Extended (q+1,2)- \\
     & Algorithm \ref{alg:q-1-omega} & network & combination network \\ \hline
  \multirow{2}{*}{$\omega$} & \multirow{2}{*}{$3$} & \multirow{2}{*}{$2$} & \multirow{2}{*}{$3$} \\
  & & & \\ \hline
 \# of & \multirow{2}{*}{$\frac{1}{2}q^2 - \frac{1}{2}q + 1$} & \multirow{2}{*}{$\frac{1}{2}q^2 + \frac{1}{2}q + 1$} & \multirow{2}{*}{$\frac{1}{2}q^2 + \frac{1}{2}q + 1$}\\
 receivers & & & \\ \hline
 \# of & \multirow{2}{*}{$\frac{1}{2}q^2 + \frac{1}{2}q + 8$} & \multirow{2}{*}{$\frac{1}{2}q^2 + \frac{3}{2}q + 3$} & \multirow{2}{*}{$\frac{1}{2}q^2 + \frac{3}{2}q + 4$} \\
 nodes & & & \\ \hline
 \# of & \multirow{2}{*}{$\frac{3}{2}q^2 - \frac{1}{2}q + 12$} & \multirow{2}{*}{$q^2 + 2q + 3$} & \multirow{2}{*}{$\frac{3}{2}q^2 + \frac{5}{2}q + 4$} \\
 edges & & & \\ \hline
\end{tabular}
\end{table}

\vspace{5pt}

\section{Summary}\label{Sec:Summary}
In this paper, we propose a particular class of multicast networks $\mathcal{N}_{\omega,\mathbf{d}}$ with topological parameters $\omega, \mathbf{d}$. By deriving a generalized Cauchy-Davenport theorem over the additive group $\mathbb{Z}_{n}$, we obtain an explicit formula on the linear solvability of $\mathcal{N}_{\omega,\mathbf{d}}$ over a base field GF($q$), which connects $\omega$ and $\mathbf{d}$ with the associated coset numbers of a multiplicative subgroup in GF($q$), rather than the conventional algebraic identity field size. Stemming from the special linear solvability behavior of $\mathcal{N}_{\omega, \mathbf{d}}$, we further formulate a subgroup order criterion for a pair of finite fields. For every pair (GF($q$), GF($q'$)) subject to the subgroup order criterion, an instance in $\mathcal{N}_{\omega,\mathbf{d}}$ can be found to be linearly solvable over GF($q$) but not over GF($q'$). Subsequently, different classes of infinitely many instances in $\mathcal{N}_{\omega, \mathbf{d}}$ are established with the special property $q_{min} < q^*_{max}$, where $q_{min}$ is the minimum field size for the existence of a linear solution and $q^*_{max}$ is the maximum field size for the non-existence of a linear solution. Moreover, it is proved that the gap $q^*_{max} - q_{min} > 0$ can tend to infinity.

Our findings suggest a new ``matching'' between the algebraic structure of a base field and the topological structure of a particular class of multicast networks, and this matching condition is both necessary and sufficient for the existence of a linear solution to the multicast networks. For a more general multicast network coding problem, it is interesting to explore and characterize similar matching conditions that are sufficient for the existence and nonexistence of a linear solution.

\vspace{3pt}

\appendices
\section*{Appendix. Theorem and Lemma Proofs}
\subsection{Proof Sketch of Lemma \ref{lemma:general_network_linear_solvability}} \label{Appendix_general_linear_solvability}
This lemma can be proven in a similar way as that of Theorem 8 in \cite{Sun_TIT}. Here we outline the sketch of the proof.
Consider an LNC with all coding coefficients being indeterminates. Without loss of generality, assume that the coding
coefficients are set to $1$ for all those adjacent pairs $(e_1, e_2)$ where $e_1$ is the unique incoming edge to some node. Assume the coding vector for the unique incoming edge to node $u_j$ is equal to the $j^{th}$ $\omega$-dimensional unit vector, and denote by $e_{jk}$ the $k^{th}$ outgoing edge from node $u_j$, $1 \leq j \leq \omega$, $1 \leq k \leq d_j$. Then, the juxtaposition of coding vectors for edges $e_{jk}$ can be represented as
\begin{equation*}
\mathbf{M'} = \left[
  \begin{array}{ccccccccccccc}
    x_{11}' & \cdots & x_{1d_1}' & 0 & \cdots & 0 &  & 0 & \cdots & 0 & x_{\omega1} & \cdots  & x_{\omega d_\omega} \\
    x_{11} & \cdots & x_{1d_1} & x_{21}' & \cdots & x_{2d_2}' &  & \vdots & \ddots & \vdots & 0 & \cdots  & 0\\
    0 & \cdots & 0 & x_{21} & \cdots & x_{2d_2} &  & 0 & \cdots & 0 & 0 & \cdots & 0\\
    0 & \cdots & 0 & 0 & \cdots & 0 & \ddots & 0 & \cdots & 0 & \vdots & \ddots & \vdots \\
    \vdots & \ddots & \vdots & \vdots & \ddots & \vdots & & x_{(\omega-1)1}' & \cdots & x_{(\omega-1)d_{\omega-1}}' & 0 & \cdots & 0 \\
    0 & \cdots & 0 &  0 & \cdots & 0 &  & x_{(\omega-1)1} & \cdots & x_{(\omega-1)d_{\omega-1}} & x_{\omega 1}' & \cdots & x_{\omega d_\omega}' \\
  \end{array}
\right]
\end{equation*}
where $x_{jk}$, $x_{jk}'$ represent coding coefficient indeterminates. Since in $\mathcal{N}_{\omega,\mathbf{d}}$ there is a receiver connected from \emph{every} set $N$ of $\omega$ layer-4 (grey) nodes with $maxflow(N) = \omega$, there exists a linear solution over GF($q$) if and only if there exists a matrix completion of $\mathbf{M'}$ over GF($q$).

Note that the coding vectors for $e_{11}, e_{21}, \cdots, e_{(\omega-1)1},e_{(\omega-1)2}$ form a full rank matrix $\left[\begin{smallmatrix}
x_{11}' & 0 & & 0 & 0 \\
x_{11} & x_{21}' & & \cdots & \cdots \\
0 & x_{21} & \cdots & 0 & 0 \\
\cdots & \cdots & & x_{(\omega-1)1}' & x_{(\omega-1)2}' \\
0 & 0 & & x_{(\omega-1)1} & x_{(\omega-1)2}
\end{smallmatrix}\right]$, %
where $x_{11}'$ is the only nonzero entry in the first row. Thus, for every matrix completion of $\mathbf{M'}$ over GF($q$), $x_{11}'$ must be set to a nonzero element in GF($q$). By a similar argument, all indeterminates $x_{jk}$ and $x_{jk}'$ must be assigned nonzero elements in GF($q'$) too. It can be subsequently seen that there is a matrix completion of $\mathbf{M'}$ over GF($q$) if and only if there is a matrix completion of matrix $\mathbf{M}$ in (\ref{eqn:general_matrix_M_for_linear_solvability_analysis}) over GF($q$), where $x_{jk}, x_{jk'}$ must be assigned two distinct nonzero elements in GF($q$) for all $1 \leq j \leq \omega$, $1 \leq k < k' \leq d_j$. Last, since coding vectors for $e_{1j_1}, e_{2j_2}, \cdots, e_{\omega j_\omega}$, $1 \leq j_k \leq d_k$, form a full rank matrix $\left[\begin{smallmatrix}
1 & 0 & & 0 & x_{\omega j_\omega} \\
x_{1j_1} & 1 & & \cdots & 0 \\
0 & x_{2j_2} & \cdots & 0 & \cdots \\
\cdots & \cdots & & 1 & 0 \\
0 & 0 & & x_{(\omega-1)j_{\omega-1}} & 1
\end{smallmatrix}\right]$,
for every matrix completion of the matrix $\mathbf{M}$ in (\ref{eqn:general_matrix_M_for_linear_solvability_analysis}) over GF($q$), $x_{1j_1}, \cdots, x_{\omega j_\omega}$ must be such assigned that $1 + (-1)^{\omega-1} x_{1j_1}x_{2j_2}\cdots x_{\omega j_\omega} \neq 0$. There is thus no difficulty to establish the equivalence between condition b) and c) in the lemma. \hfill $\blacksquare$

\subsection{Proof of Theorem \ref{thm:Cauchy-Davenport-Generalization}} \label{Appendix_Cauchy-Davenport}
First, note that if the elements in $A_1$ are in a same coset of the stabilizer group $H$, then $|A_1 + H| = |H|$. Moreover, if they are in $l$ different cosets of $H$, then $|A_1 + H| = l|H|$. Since the elements in each $A_j$, $1 \leq j \leq k$, belong to  at least $\left\lceil\frac{|A_j|}{|H|}\right\rceil$ different cosets of $H$,
\begin{equation}
\label{eqn:Appendix_Stablizer}
|A_j + H| \geq |H|\left\lceil\frac{|A_j|}{|H|}\right\rceil,~~\forall 1 \leq j \leq k.
\end{equation}
Subsequently, it suffices to prove that
\begin{equation}
\label{eqn:Appendix_proof_1}
|A_1 + A_2 + \cdots + A_k| \geq |A_1 + H| + |A_2 + H| + \cdots + |A_k + H| - (k-1)|H|,
\end{equation}
because this, together with (\ref{eqn:Appendix_Stablizer}) implies (\ref{eqn:Kneser_Generalization}).

The Kneser's Additive Theorem (See, e.g., Theorem 3.3.2 in \cite{Additive_Number_Theorem}) asserts that for any two nonempty finite subsets $A$ and $B$ in an abelian group $G$, the inequality $|A + B| \geq |A+H'| + |B+H'| - |H'|$ holds, where $H'$ is the stabilizer group of $A+B$. Thus, the case of $k = 2$ for inequality (\ref{eqn:Appendix_proof_1}) degenerates to the Kneser's Theorem. By induction on $k$, we may assume that the theorem holds when $k$ is substituted by $j = k-1$. Thus,
\begin{equation}
\label{eqn:Appendix_proof_2}
|A_1 + \cdots + A_j + A_{j+1}| \geq |A_1 + H| + \cdots + |A_{j-1} + H| + |A_j + A_{j+1} + H| - (j-1)|H|.
\end{equation}

Since $H$ is a group, $A_j + A_{j+1} + H = (A_j + H) + (A_{j+1} + H)$. By applying the Kneser's Theorem again, we have
\begin{equation}
\label{eqn:Appendix_proof_3}
|A_j + A_{j+1} + H| \geq |A_j + H + H_1| + |A_{j+1} + H + H_1| - |H_1|,
\end{equation}
where $H_1$ is the stabilizer group of $A_j + A_{j+1} + H$ in $\mathbb{Z}_n$. Since $A_j + A_{j+1} + H + H_1 = A_j + A_{j+1} + H$, $H$ is contained in the stabilizer group $H_1$. On the other hand, because
\[
A_1 + \cdots + A_j + A_{j+1} + H + H_1 = A_1 + \cdots + A_j + A_{j+1} + H = A_1 + \cdots + A_j + A_{j+1},
\]
where the first equality holds for $H_1$ is the stabilizer group of $A_j + A_{j+1} + H$ and the second equality holds for $H$ is the stabilizer group of $A_1 + \cdots +A_{j} + A_{j+1}$, $H + H_1 = H_1$ is contained in the stabilizer group $H$ too. Thus, $H = H_1$, and (\ref{eqn:Appendix_proof_3}) becomes
\[
|A_j + A_{j+1} + H| \geq |A_j + H| + |A_{j+1} + H| - |H|.
\]
This, together with (\ref{eqn:Appendix_proof_2}) implies (\ref{eqn:Appendix_proof_1}) as desired, so condition (\ref{eqn:Kneser_Generalization}) holds. Since the order of the stabilizer group $H$ divides $n$, inequality (\ref{eqn:Cauchy-Davenport-Generalization}) is a direct consequence of (\ref{eqn:Kneser_Generalization}). \hfill $\blacksquare$

\vspace{3pt}

\subsection{Proof of Lemma \ref{lemma:Diophantine_approximation}} \label{Appendix_Diophantine}
Let $\delta_0$ be an arbitrary real number with $0 < \delta_0 < c_1 - c_2$. If $\frac{\delta}{n_1^{\delta_0} - 1} < 1$, then set $k_0 = 0$. Otherwise, set $k_0 = \left\lceil\log_{n_2} \frac{\delta}{n_1^{\delta_0} - 1}\right\rceil$. When $k' > k_0$, we have $n_2^{k'} > n_2^{k_0} \geq \frac{\delta}{n_1^{\delta_0} - 1}$, and hence $n_2^{k'}n_1^{\delta_0} > n_2^{k'} + \delta$, that is,
\[
\log_{n_1}n_2^{k'} + \delta_0 > \log_{n_1}\left(n_2^{k'} + \delta\right).
\]
Consequently, it suffices to show that there are infinitely many positive integers $k$, $k'$ such that
\begin{equation}
\label{eqn:Appendix_B_proof}
\log_{n_1}n_2^{k'} + k_0 + c_1 > k > \log_{n_1}n_2^{k'} + k_0 + \delta_0 + c_2.
\end{equation}
If $c_1 > 1 + \delta_0 + c_2$, then for every $k' \geq 1$, there is an integer $k$ in the interval $(\log_{n_1}n_2^{k'} + k_0 + \delta_0 + c_2, \log_{n_1}n_2^{k'} + k_0 + c_1)$. Assume that $c_1 \leq 1 + \delta_0 + c_2$. Since $n_1$, $n_2$ are coprime, $\log_{n_1}n_2$ is an irrational number. Thus, according to the Weyl's equidistribution theorem (See, e.g., \cite{Book_for_Equidistribution}), $(\log_{n_1}n_2)\mathbb{Z}/\mathbb{Z}$ is uniformly distributed over $\mathbb{R}/\mathbb{Z}$, where $\mathbb{R}$ is the real number field. Therefore, a fraction $c_1 - c_2 - \delta_0$ of all positive integers $k'$ can make the interval $(\log_{n_1}n_2^{k'} + k_0 + \delta_0 + c_2, \log_{n_1}n_2^{k'} + k_0 + c_1)$ contain an integer, that is, to make the inequality (\ref{eqn:Appendix_B_proof}) hold. \hfill $\blacksquare$

\vspace{3pt}

\subsection{Proof of Theorem \ref{thm:p_vs_p'_general_odd}} \label{Appendix_General_p_vs_p}
Let $q$ be a prime power $p^j$ such that $q - 1$ has a divisor $d$ larger than 2. Thus, $q^k - 1$ is divisible by $d$ for all $k \geq 1$. Let $a$ be the smallest integer such that $p'^{a} - 1$ is divisible by $d$, if there does not exist such one, then set $a = 1$. Thus, $p'^{ak' + 1} - 1$ is not divisible by $d$ for all $k' \geq 1$.

Let $k, k'$ be arbitrary positive integers subject to
\begin{equation}
\label{eqn:inequality_p_vs_p'_general_odd}
\log_q(p'^a)^{k'} + \log_q{p'} > k > \log_q\left((p'^a)^{k'} + \frac{1}{p'(d^2 -2d)}\right) + \log_q\left(p' - \frac{p'}{(d-1)^2}\right).
\end{equation}
According to Lemma \ref{lemma:Diophantine_approximation}, there are infinitely many such choices of $k$ and $k'$. We next show that the subgroup order criterion in Definition \ref{Def:Subgroup_Order_Criterion} holds for (GF($q^k$), GF($p'^{ak'+1}$)) by setting $G$ to be the subgroup in GF($q^k$)$^\times$ of order $\frac{q^k-1}{d}$.

In the case that $p'^{ak'+1} - 1$ does not have a proper divisor no smaller than $\frac{q^k - 1}{d}$, the subgroup order criterion naturally holds for (GF($q^k$), GF($p'^{ak'+1}$)) as desired. Otherwise, let $d'$ be an arbitrary proper divisor of $p'^{ak'+1} - 1$ no smaller than $\frac{q^k - 1}{d}$. By the second inequality in (\ref{eqn:inequality_p_vs_p'_general_odd}), it can be deduced that
\begin{equation}
\label{eqn:inequality_p_vs_p'_general_odd1}
\frac{d-1}{d}\left(q^k - 1\right) > \frac{d-2}{d-1}\left(p'^{ak'+1}-1\right).
\end{equation}
Since the divisor $d$ of $q-1$ is selected larger than 2, $\frac{d-2}{(d-1)^2} \geq \frac{1}{d+1}$ and consequently,
\[
d' \geq \frac{q^k-1}{d} > \frac{d-2}{(d-1)^2}\left(p'^{ak'+1}-1\right) \geq \frac{1}{d+1}\left(p'^{ak'+1}-1\right).
\]
Moreover, since $d$ does not divide $p'^{ak'+1}-1$,
\begin{equation}
\label{eqn:inequality_p_vs_p'_general_odd2}
d' \geq \frac{1}{d-1}\left(p'^{ak'+1}-1\right).
\end{equation}
By combining (\ref{eqn:inequality_p_vs_p'_general_odd1}) and (\ref{eqn:inequality_p_vs_p'_general_odd2}), we obtain
\[
q^k - \frac{q^k-1}{d} - 1 > \frac{d-2}{d-1}\left(p'^{ak'+1}-1\right) \geq p'^{ak'+1} - d' - 1.
\]
This implies that the subgroup order criterion in Definition \ref{Def:Subgroup_Order_Criterion} indeed holds for (GF($q^k$), GF($p'^{ak'+1}$)) as desired. Consider the network $\mathcal{N}_{\omega,\mathbf{d}}$ with $\omega$ set to satisfy (\ref{eqn:omega_sharp_bound}) and $\mathbf{d}$ to be the $\omega$-tuple $(\frac{q^k-1}{d}, \cdots, \frac{q^k-1}{d}, \frac{d-1}{d}(q^k-1))$. Theorem \ref{Thm:Implication_Subgroup_Order_Criterion} then shows that $\mathcal{N}_{\omega,\mathbf{d}}$ is linearly solvable over GF($q_{min}$) with $q_{min} = q^k$ but not linearly solvable over GF($p'^{ak'+1}$), where $p'^{ak'+1} > q^k$ is due to the first inequality in (\ref{eqn:inequality_p_vs_p'_general_odd}). \hfill $\blacksquare$

\end{document}